\documentclass[12pt]{elsart}
\usepackage{color}
\usepackage{graphicx}
\usepackage{amssymb}
\usepackage{amsfonts}
\usepackage{amsmath}

\def\lsim{\stackrel{\scriptstyle <}{\phantom{}_{\sim}}}
\def\gsim{\stackrel{\scriptstyle >}{\phantom{}_{\sim}}}

\begin{document}
\begin{frontmatter}

\title{Quark scattering off quarks and hadrons\\
}
\author[JINR]{A.V~Friesen},
\author[JINR]{Yu.L.~Kalinovsky}
\and \author[JINR]{V.D.~Toneev}
\address[JINR]{Joint Institute for Nuclear Research,
 141980 Dubna, Moscow Region, Russia}

\begin{abstract}The in-medium elastic scattering $qq\to qq,
q\bar{q}\to q\bar{q}$ and $\bar{q}\bar{q}\to \bar{q}\bar{q}$ is
calculated within the two-flavor Polyakov-loop-extended
Nambu-Jona-Lasinio model. The integral and differential
quark-quark scattering, its energy and temperature dependence are
considered and their flavor dependence is emphasized. The
comparison  with results of other approaches is presented. The
consideration is implemented to the case of quark-pion scattering
characterizing the interaction between quarks and hadrons
in a kinetic multiphase treatment, and the first estimate of the
quark-pion cross sections is given. A possible application of the
obtained results to heavy ion collisions is shortly discussed.
\end{abstract}
\end{frontmatter}

\section{Introduction}

To describe high-energy nuclear physics, the knowledge of the
in-medium behavior of quasiparticles   such as  quarks, gluons,
mesons and baryons including their antiparticles is crucial. The
quantum chromodynamics (QCD)  seems to be the best tool to proceed
to this study. Nevertheless, it is well known that the direct
implementation of the QCD Lagrangian is not feasible in this state
except for some particular cases. In order to avoid the QCD
difficulties, some effective models were developed.

 In this respect, the low-energy particle sector is well described by
effective chiral theories of QCD, the Nambu and Jona-Lasinio (NJL)
model~\cite{NJL61}. The advantage of this model is  that it can be
studied in the entire temperature range. The NJL model also offers
a simple intuitive view of chiral symmetry breakdown and
restoration via the realization of the quark-antiquark pairing
similar to the BCS theory of superconductivity. However, a simple
point-like interaction form of the model does not ensure its
renomalizability, and a cutoff scale $\Lambda$ must be introduced
in the theory. The impossibility to treat the
confinement-deconfinement phase transition and the absence of
gluons are other important defects of the NJL model.

To eliminate partially these defects, it has recently been
proposed to couple the quarks to a Polyakov loop~\cite{Po78,Fu04}
as a mechanism that could simulate the confinement, even if the
model does not consider the color degrees of freedom as done in
QCD. This realized approach is called the
Polyakov-Nambu--Jona-Lasinio model
(PNJL)~\cite{HF03,RTW06,RRTW07,CRSH10,MMR07}. Some recent results
show that this approach,  exhibiting a smooth crossover at zero
baryon density and a first-order phase transition at a large baryon
chemical potential,   provides some
advantages~\cite{HF03,CRSH10}. In particular, the extended model
allows one to correctly reproduce lattice data of QCD
thermodynamics~\cite{RTW06,SSKY10} as well as to improve the NJL
model at low temperature due to the suppression of the
contribution of colored states. In addition, the PNJL model is
more efficient for describing the restoration of the chiral
symmetry by a rapid decrease in the effective masses of the
quarks~\cite{SSKY10}. Nevertheless, the phase structure and
its dependence on thermodynamic variables is still an open problem
and raises some interesting questions including chiral symmetry
restoration, color superconductivity, and charged pion
condensation  phenomena. In particular, it has been found that if
the isospin chemical potential $\mu_I$ in a charge neutral quark
matter is  lower than the critical value required for the
realization of the pion condensation,  pions do not condense and,
therefore, even above the critical temperature a bound state with
the pion quantum numbers can be formed~\cite{ACG08}.

At high temperatures ($T>$ 300 MeV) it is supposed that the PNJL
description is reliable up to a temperature of approximately
2.5$T_c$. For still higher temperatures the transverse gluons,
ignored in the PNJL treatment, are expected to be
non-negligible~\cite{MOM04}. The strong interaction in such a
nonperturbative regime of the deconfinement phase is taken into
account through an effective temperature-dependent mass for the
gluons with a Polyakov-loop background, leaving open the
possibility that lighter quasiparticles propagate in the
medium~\cite{RAC12,SR12,OCG13}.

Allowability of quark-gluon degrees of freedom along with hadronic
ones means that the model for heavy-ion collisions should be
multiphase in nature and include possible phase transitions
between different phases. Generally,  this complicated situation can
be described in terms of hydrodynamics or kinetics which have
their own advantages and disadvantages. The use of kinetics for
the quark-gluon phase needs knowledge of in-medium cross sections
for its constituents. In A MultiPhase Transport (AMPT)
model~\cite{AMPT} this phase is described in terms of the parton
cascade with the partonic elastic cross sections estimated within
the perturbative QCD (pQCD). The effect of the surrounding matter
was roughly included by introducing the effective Debye mass.

The microscopic  quark dynamics is studied in a more elaborated
way in the Parton Hadron String Dynamics (PHSD)
model~\cite{Ca09,PHSD,BCKL11} where the plasma evolution is solved
by a Kadanoff-Baym type equation. Here the potentials between the
plasma constituents are chosen in such a way that the model
equation of state is consistent with lattice calculations. The
cross sections are derived from the spacelike part of the
interaction and are employed for the scattering interactions among
the plasma constituents. In this model, gluons as well as quarks
acquire a large mass when approaching the phase transition.
Therefore, the prehadrons which are created in the phase
transition are rather heavy.  Another model which allows for these
studies is a gluonic cascade realized in the Boltzmann Approach to
Multi Parton Scattering (BAMPS)~\cite{UFXG10}. The gluon emission
and interaction during the expansion stage  of the QGP move the
system towards equilibrium. A parton cascade approach with a
pQCD inspired cross section was applied also at the RHIC energy to
study scaling properties of the elliptic flow~\cite{FC09} and
'chemical' composition of the quark-gluon plasma~\cite{SC12}.

The quantum molecular dynamics of the expanding $q/\bar{q}$ plasma
has been proposed  recently~\cite{MA12}. Properties of quarks as
well as elastic scattering cross sections were calculated within
the three-flavor NJL model.

All the kinetic multiphase  models mentioned above should describe
a transition from one to another phase: from quarks-gluons to
hadrons, in our case. This smooth transition is simulated by a
possible coalescence of quark-antiquark or three quarks, being
close to each other in coordinate and momentum space, to form a
meson or a baryon thereby creating a mixed parton-hadron phase
which is a typical feature of the crossover phase transition.
 As was demonstrated in Ref.~\cite{SYY93} in terms of a simple
thermodynamically consistent statistical model, the gluon-glueball
system exhibiting a crossover phase transition shows the first
order phase transition if the interaction between mixed phase
constituents is neglected. As to chiral NJL-like models, the
appearance of a first-order chiral phase transition is a
characteristic feature of the simplest versions of these models.
Sensitivity of the phase structure to the parameters
characterizing the quasiparticle interaction has noted many years
ago. In particular, it was shown that the  location of the QCD
critical point moves in accord with the repulsive vector-channel
interaction which may result in the disappearance of the critical
end-point at sufficiently large values of the vector interaction
coupling turning it into crossover~\cite{AY89,Fu08,CNB10,BHW04}. A
pronounced impact on the phase diagram is also given by breaking
of the $U(1)$ symmetry due to the axial anomaly in QCD which is
introduced into models by adding the Kobayashi-Maskava-'t Hooft
interaction to be responsible for the large mass of the $\eta'$
meson. The decrease of this interaction strength may dismiss the
associated existence of a critical point in the phase
diagram~\cite{Fu08,BHW04}.

In this work, we want to make a step towards account for the
interaction  between constituents of  quark-gluon and hadronic phases.
Basing on the PNJL model we give here the first estimate for interaction
of quarks/antiquarks with pions which is expected to be a dominant
component of this type of interactions.

Thus, the purpose of this paper is to calculate quark-quark,
quark-antiquark and antiquark-antiquark  cross sections and
generalize this approach to the case of quark-pion scattering. The
consideration is based on the chiral two-flavor PNJL model the key
points of which are remind in the next Section II. In Sections III
and IV, the main equations are given for different channels of
quark-(anti)quark elastic scattering and are generalized to the
quark-hadron case in Section IV. Their numerical results for the
$qq$ and $qH$ processes are presented and discussed in Section V.
We conclude the obtained results in Section VI.

\section{The PNJL model used}

The deconfinement in a pure $SU_f(2)$ gauge theory can be
simulated by introducing  a complex Polyakov loop field. The
two-flavor PNJL model is used with the following
Lagrangian~\cite{RTW06,pnjl1,hansen}~:
\begin{eqnarray}
 \label{Lpnjl}
\mathcal{L}_{\rm PNJL}=\bar{q}\left(i\gamma_{\mu}D^{\mu}-\hat{m}_0
\right) q+ G  \left[\left(\bar{q}q\right)^2+\left(\bar{q}i\gamma_5
\vec{\tau} q \right)^2\right]
-\mathcal{U}\left(\Phi[A],\bar\Phi[A];T\right)~,
\end{eqnarray}
where scalar and pseudoscalar interactions are taken into account,
$G$ is the coupling constant, $\vec{\tau}$ is the Pauli matrix in
the flavor space, $\bar{q}$ and $q$ are the quark fields (color
and flavor indices are suppressed), $\hat{m}_0$ is the diagonal
matrix of the current quark mass, $\hat{m}_0 = \mbox{diag} \,
(m^0_u, m^0_d)$ and $m^0_u = m^0_d=m_0$. The vectorial and axial
interaction terms are neglected in Eq.~(\ref{Lpnjl}).

The quark fields are related to the gauge field $A^\mu$ through
the covariant derivative $D^\mu = \partial^\mu -iA^\mu$, where the
gauge field is $A^\mu = \delta_0^\mu A^0 = -i \delta_4^\mu A_4$
(the Polyakov calibration). The field $\Phi$ is determined by
tracing the Polyakov loop $L(\vec{x})$~\cite{RTW06}: $\Phi[A] =
\frac{1}{N_c} \mbox{Tr}_c L(\vec{x})$, where $L(\vec{x})  =
\mathcal{P} \exp \left[ \displaystyle i \int_{0}^{\beta} d \tau
A_4 (\vec{x}, \tau)  \right]$.

The gauge sector of the Lagrangian density (\ref{Lpnjl}) is
described by an effective potential
$\mathcal{U}\left(\Phi[A],\bar\Phi[A];T\right)$ fitted to lattice
QCD simulation results in a pure $SU(3)$ gauge theory at finite
$T$~\cite{RTW06,RRW07}
\begin{eqnarray}\label{effpot}
\frac{\mathcal{U}\left(\Phi,\bar\Phi;T\right)}{T^4}
&=&-\frac{b_2\left(T\right)}{2}\bar\Phi \Phi-
\frac{b_3}{6}\left(\Phi^3+ {\bar\Phi}^3\right)+
\frac{b_4}{4}\left(\bar\Phi \Phi\right)^2 \\ \label{Ueff}
b_2\left(T\right)&=&a_0+a_1\left(\frac{T_0}{T}\right)+a_2\left(\frac{T_0}{T}
\right)^2+a_3\left(\frac{T_0}{T}\right)^3~.
\end{eqnarray}
The parameters of the effective potential (\ref{effpot}) and
(\ref{Ueff}) defined by fitting to the lattice results
 are summarized in~\cite{FKT} with the model parameter value of $T_0 = 0.19$ GeV.

The grand potential for the PNJL theory in the mean-field
approximation is given by the following equation~\cite{hansen}~:
\begin{eqnarray} \label{potpnjl}
\Omega (\Phi, \bar{\Phi}, m, T, \mu) &=&
\mathcal{U}\left(\Phi,\bar\Phi;T\right) + G \langle \bar{q}q
\rangle ^2 +\Omega_q
\end{eqnarray}
where
\begin{eqnarray}
\Omega_q = -2 N_c N_f \int \frac{d^3p}{(2\pi)^3} E_p \nonumber -
2N_f T \int \frac{d^3p}{(2\pi)^3} \left[ \ln N_\Phi^+(E_p)+ \ln
N_\Phi^-(E_p) \right]~,
\end{eqnarray}
 $E_p$ is the quark energy, $E_p=\sqrt{{\bf p}^2+m^2}$,
$E_p^\pm = E_p\mp \mu$ ($\mu$ is the chemical potential) and $
N_\Phi^\pm(E_p)$ is the partition density with
\begin{eqnarray}
&& N_\Phi^+(E_p) = \left[ 1+3\left( \Phi +\bar{\Phi} e^{-\beta
E_p^+}\right) e^{-\beta E_p^+} + e^{-3\beta E_p^+}
\right]^{-1},  \nonumber \\[-0.1cm]\\[-0.25cm]
&& N_\Phi^-(E_p) = \left[ 1+3\left( \bar{\Phi} + {\Phi} e^{-\beta
E_p^-}\right) e^{-\beta E_p^-} + e^{-3\beta E_p^-}
\right]^{-1}.\nonumber
\end{eqnarray}

The gap equation for the constituent quark mass is obtained by
solving the equation ${\partial \Omega (\Phi, \bar{\Phi}, m, T,
\mu) }/{\partial m}=0 $ with the grand potential (\ref{potpnjl}):
\begin{eqnarray} \label{gap-eq}
m = m_0-N_f  G <{\bar q}q> =m_0 + 8 G N_c N_f \int_{\Lambda}
\frac{d^3p}{(2\pi)^3} \frac{m}{E_p} \left[ 1 - f^+_{\Phi} - f^-
_{\Phi}\right]~,
\end{eqnarray}
 where $f_\Phi^+$, $f_\Phi^-$ are the modified Fermi functions
\begin{eqnarray}
f_\Phi^+ = ((\Phi + 2\bar{\Phi}e^{-\beta E^+})e^{-\beta E^+}
+e^{-3\beta E^+})N_\Phi^+, \nonumber\\[-0.1cm]  \label{fermi}\\[-0.25cm]
f_\Phi^- = ((\bar{\Phi} + 2{\Phi}e^{-\beta E^-})e^{-\beta E^-}
+e^{-3\beta E^-})N_\Phi^- \nonumber
\end{eqnarray}
and $\beta=1/T$ is the inverse temperature.

\begin{figure} [h]
\centerline{
\includegraphics[width = 7.5cm] {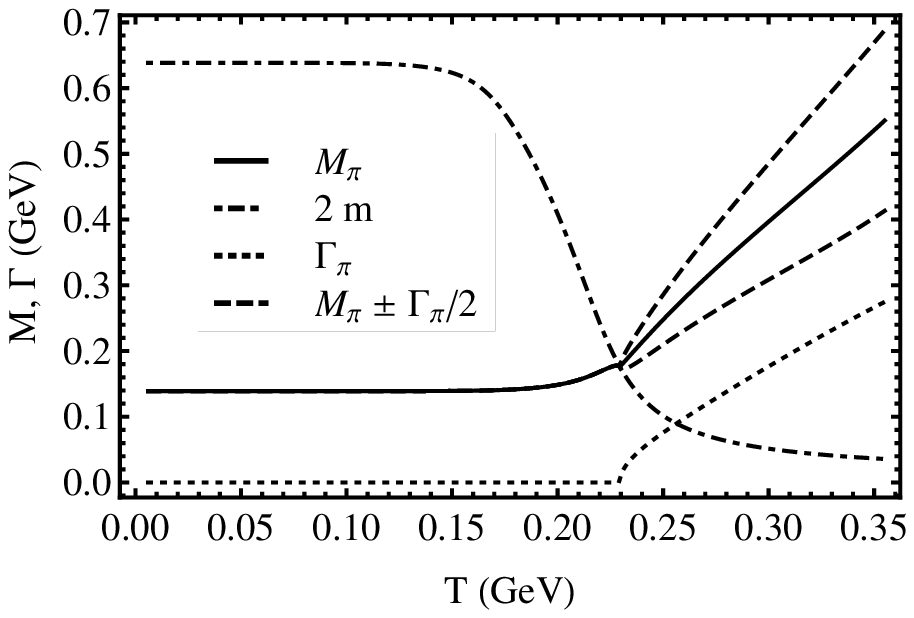}
\includegraphics[width = 7.5cm] {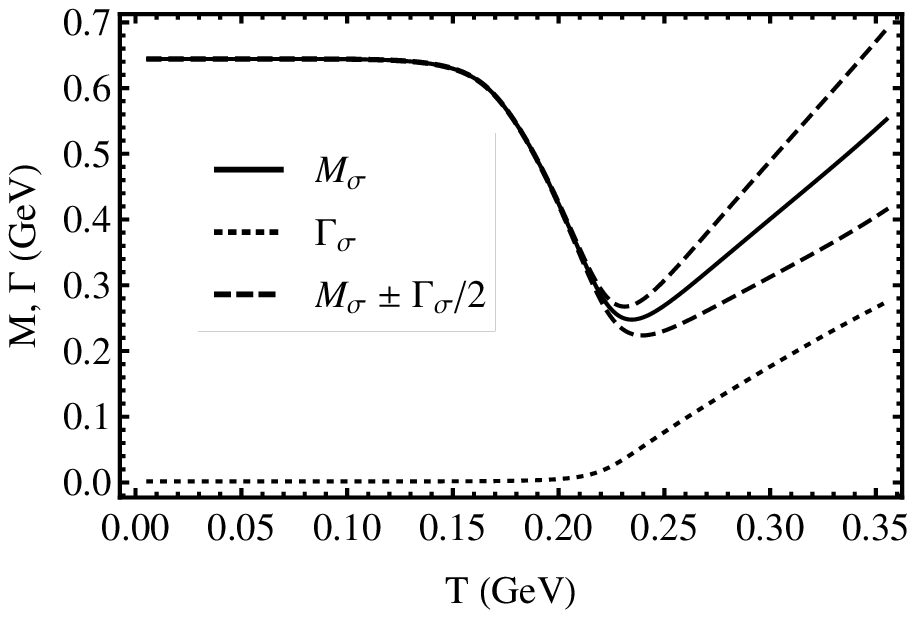}
} \caption{Meson mass $M_{\pi/\sigma}\pm \Gamma_{\pi/ \sigma}/2$,
the double quark mass and the meson width $\Gamma_{\pi/ \sigma}$
in the PNJL model.} \label{spectra}
\end{figure}

In our model, the mesons as  $q-\bar{q}$ bound states  are constructed by
the quark-antiquark interaction within the random-phase approximation
\cite{Kl92}. It leads to the explicit form for the meson
propagators

\begin{eqnarray}\label{Dps}
D_{\pi/\sigma}(k_0, {\bf k}) =
\dfrac{G}{1-2G\Pi_{\pi/\sigma}(k_0, {\bf k})}.
\end{eqnarray}
In the self-consistent Hartree limit masses of  bound states are
determined as poles of the meson propagators in the random-phase
approximation. Therefore, the $\pi$ and $\sigma$ meson masses are
the solutions  of the equation
\begin{equation}\label{eq8}
1 - 2G \ \Pi_{{\pi/\sigma}}(k_0,{\bf k}) = 0,
\end{equation}
where $k^2 = M^2_\pi$ and $k^2 = M^2_\sigma$ in pseudoscalar and
scalar sectors, respectively, and $\Pi_{{\pi}/{\sigma}}$ are the
correlation functions  with the quark propagator
$S_{\pi/\sigma}(k)$~\cite{Kl92}
\begin{eqnarray}
&& i\Pi_\pi (k^2) = \int \frac{d^4p}{(2\pi)^4} \ \mbox{Tr}\,
\left[ i \gamma_5 \tau^a S_\pi(p+k) i \gamma_5 \tau^b S_\pi(p)
 \right],  \label{Polpi} \\
 && i\Pi_\sigma  (k^2)= \int \frac{d^4p}{(2\pi)^4} \ \mbox{Tr}\, \left[
 i S_\sigma(p+k)i S_\sigma(p)
 \right].
  \label{Polsig}
\end{eqnarray}

Equation~(\ref{eq8}) means that  in the pole approximation
to be reasonable in the $1/N_c$ consideration, the  meson propagators
$D_{\pi/\sigma}$ can be written as
\begin{eqnarray}
D_{\pi/\sigma}(k_0,{\bf k}) =
\dfrac{G}{1-2G\Pi_{\pi/\sigma}(k_0,{\bf k})} \approx
\dfrac{g^2_{\pi q\bar{q}/\sigma q \bar{q}}}{k^2-M_{\pi/\sigma}^2}.
\end{eqnarray}
As in Refs.~\cite{Kl92}-\cite{RKH96PC}, both the pion-quark $g_{\pi q\bar{q}}(T,\mu)$
and sigma-quark $g_{\sigma q\bar{q}}(T,\mu)$
coupling strengths can be obtained now from
$\Pi_{{\pi}/{\sigma}}$  :
\begin{eqnarray}
g_{\pi q\bar{q} /\sigma q\bar{q}}^{-2}(T, \mu) = \frac{\partial\Pi_{{\pi}/{\sigma}}(k^2)} {\partial k^2}\vert_{^{k^2 = m_\pi^2}_{k^2
=m_\sigma^2} }. \label{couple}
\end{eqnarray}

 The regularization parameter $\Lambda$,  the
quark current mass $m_0$, the coupling strength G, the parameter
$T_0$ and the Mott temperature in these calculations are presented
in Table~\ref{param}. In the PNJL model there are two critical
temperatures: The critical temperature of the chiral transition
and the deconfinement temperature  which can coincide for some
model parameter set~\cite{FKT}. The first quantity is obtained as
a minimum of $\partial m/\partial T$ and the second one as a
maximum of $\partial \Phi/\partial T$. For the vanishing chemical
potential the temperature of the chiral transition is $T_c=$0.210
GeV for our model parameters.

\begin{table}[h]
\caption{The model parameters}
\begin{center}
\begin{tabular}{ccccc}
\hline
$m_0$ [MeV] & $\Lambda$ [GeV] & $G$ [GeV]$^{-2}$ & $T_0$ [GeV] & $T_{\rm Mott}$  [GeV]  \\
\hline
5.5 & 0.639 & 5.227 &  0.19 & 0.231 \\
\hline
\end{tabular}
 \label{param}
 \end{center}
 \end{table}

The temperature dependence of $\pi$ and $\sigma$-meson masses is
shown in Fig.~\ref{spectra},  its behavior is typical for NJL-like
models. The double mass of constituent quarks $2m$ is also plotted
in this figure. As is seen, the curve for $2m$ crosses the
$\pi$-meson one at $T=$0.231 GeV allowing the pion to dissociate
into their constituents. This so-called Mott temperature  is
 sometimes considered as a "soft" form of deconfinement. Note that
$T_{Mott}>T_c$. At higher temperatures both scalar and
pseudoscalar meson masses jointly increase. The decay width of
$\pi$ and $\sigma$ mesons, $\Gamma_\pi$ and $\Gamma_\sigma$,
monotonically grows above the Mott temperature as $T$ increases.

\begin{figure} [h]
\centerline{
\includegraphics[width = 7.5cm] {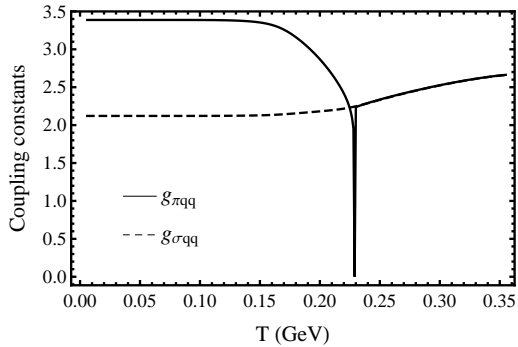}}
\caption{Coupling constants $g_{\sigma qq}$ (dashed line) and
$g_{\pi qq}$ (solid line).} \label{cc}
\end{figure}

As demonstrated in Fig.~\ref{cc} and in accordance with
Eq.~(\ref{couple}), the quark-$\sigma$ coupling constant is rather
weakly sensitive to the temperature slowly increasing with $T$ but
in the quark-pion   case the coupling constant exhibits a kink
singularity just at the Mott temperature. Technically, this
feature results in the coupling strengths approaching zero for
$T\to T_{Mott}$ from below. This behavior differs markedly from
the behavior of the couplings when evaluated in the chiral limit.

\section{Scattering cross section for the $ qq \rightarrow qq $ and $ q \overline{q}
\rightarrow q \overline{q} $ processes}

\subsection*{Quark-quark elastic scattering}
\begin{figure} [h]
\centerline{
\includegraphics[width=4.cm]{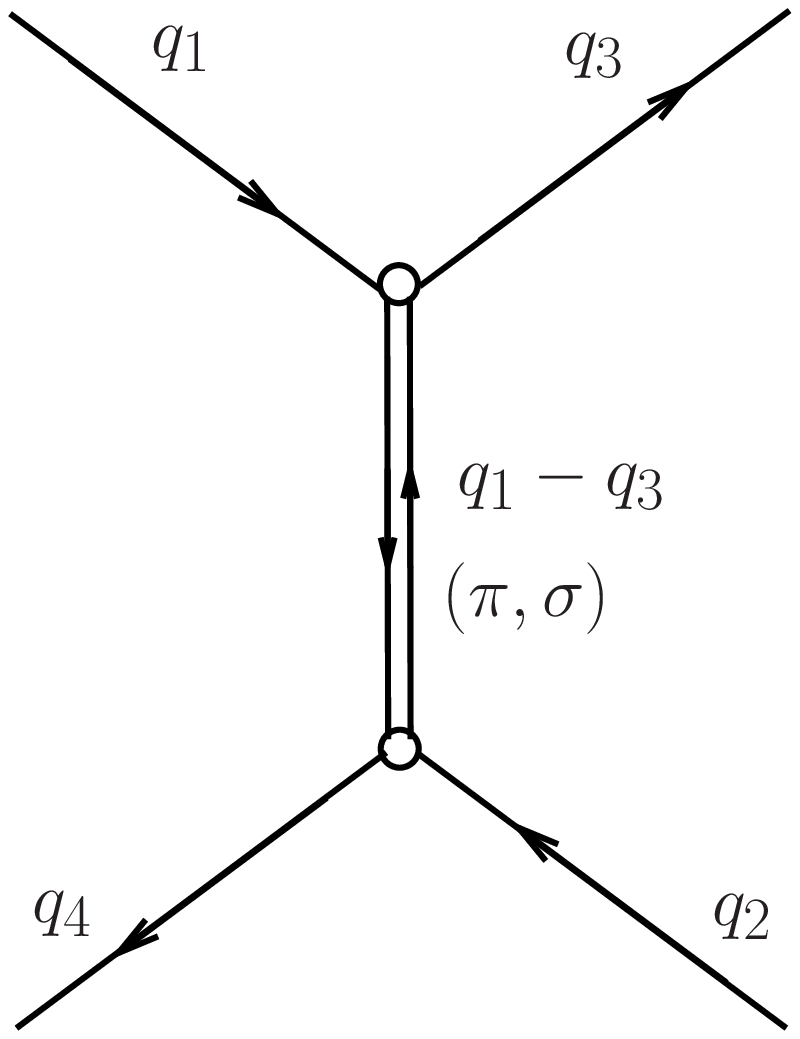}
\hspace{1cm}
\includegraphics[width=4.cm]{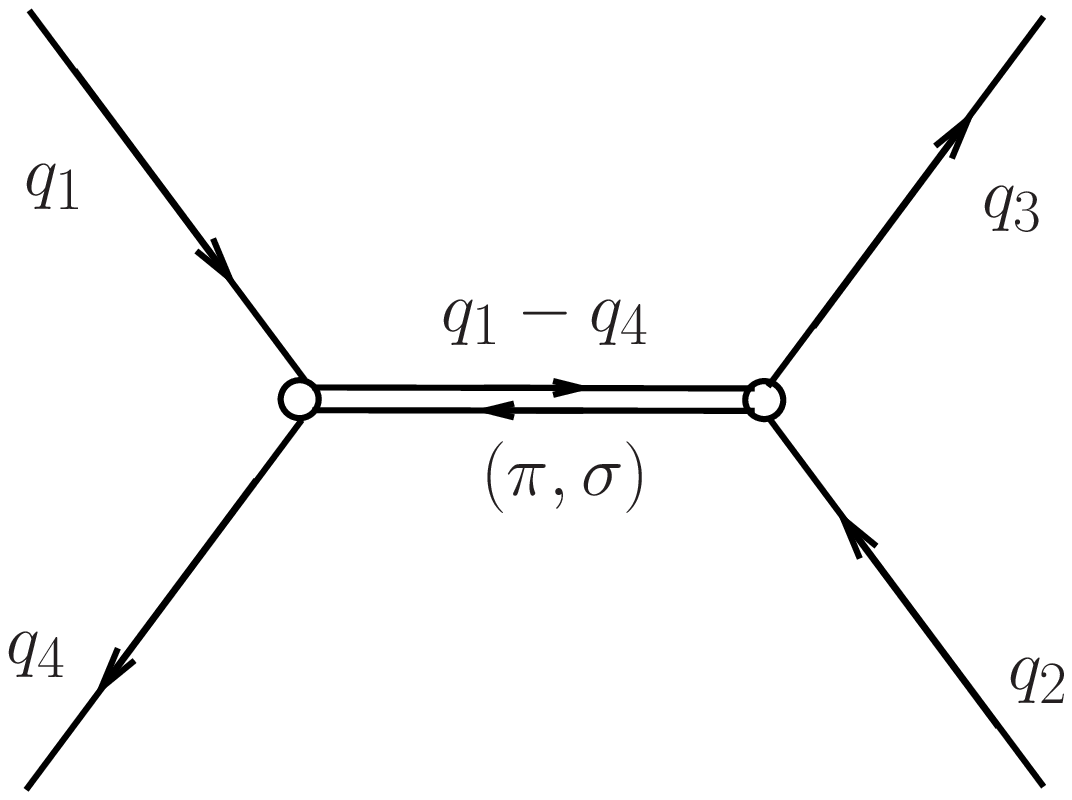}
} \caption{Diagrams of $qq \rightarrow qq$ for $t$-channel (left)
and $u$-channel (right)}
 \label{fig_diagqq}
\end{figure}

Let us consider now the quark scattering processes. The amplitude
of the quark-quark scattering to $1/N_c$ order is given by two
diagrams presented in Fig.~\ref{fig_diagqq} with taking into
account the channels with pion and $\sigma$-meson creation in an
intermediate state:

\begin{eqnarray}
-i T_t &=& \overline{u}(q_3)\Gamma_\pi u(q_1)\frac{1}{(q_1-q_3)^2 -
M_\pi^2}\overline{u}(q_4)\Gamma_\pi u(q_2) +  \nonumber \\
&& \overline{u}(q_3)\Gamma_\sigma u(q_1)\frac{1}{(q_1-q_3)^2 - M_\sigma^2}\overline{u}(q_4)\Gamma_\sigma u(q_2), \\
-i T_u &=& \overline{u}(q_4)\Gamma_\pi u(q_1)\frac{1}{(q_1-q_4)^2 -
M_\pi^2}\overline{u}(q_3)\Gamma_\pi u(q_2) + \nonumber \\
&&\overline{u}(q_4)\Gamma_\sigma u(q_1)\frac{1}{(q_1-q_4)^2 -
M_\sigma^2}\overline{u}(q_3)\Gamma_\sigma u(q_2),
\end{eqnarray}
where $\Gamma_\pi = (i\gamma _5) \cdot g_{\pi qq}$ and
$\Gamma_\sigma = {\bf 1}\cdot g_{\sigma qq}$. Here $s$ and $t$ are
the usual Mandelstam variables. After appropriate transformations
and bearing in mind that the total quark-quark amplitude is
$\displaystyle T_{qq} = \dfrac{1}{4N_c^2} \sum_{c} |T_t + T_u|^2 $
we get the following result~:
\begin{eqnarray}
|T_{t}|^2 &=& \left( |D_t^\sigma |^2(t-4m^2)^2 + |D_t^\pi |^2 t^2 \right),\\
|T_{u}|^2 &=& \left( |D_u^\sigma |^2(u-4m^2)^2 + |D_u^\pi |^2 u^2 \right),\\
T_t T^*_u &=& -\frac{1}{2 N_c}( D_t^\sigma D_u^\sigma (tu + 4
m^2(u+t) - 16m^2)
- D_t^\sigma D_u^\pi u (t- 4m^2) \nonumber \\
& - & D_t^\pi D_u^\sigma t (u - 4m^2) + D_t^\pi D_u^\pi tu )~,
\end{eqnarray}
 where the effective meson propagators in the $t$ and $u$ channels
are
\begin{equation}
D^{\sigma/ \pi}_t = \frac{g^2_{\sigma qq/ \pi qq}}{t - M_{\sigma/ \pi}^2}, \ \ \
D^{\sigma/ \pi}_u = \frac{g^2_{\sigma qq/ \pi qq}}{u - M_{\sigma/ \pi}^2}~.
\end{equation}
with the coupling constants $g_{\sigma qq\ \pi qq}$ defined from
Eq.~(\ref{couple}).

The differential cross section has the form
\begin{equation}
\frac{d\sigma_{el}}{dt} = \frac {| T |^2} {16 \pi \lambda (s,
m^2)} \label{dif_cs}
\end{equation}
with $ \lambda(s, m^2) = (s - 4 m^2) s $.

Then the  total cross section of the $ qq \rightarrow qq $
and $ q \overline{q} \rightarrow q \overline{q} $ elastic
scattering is:
\begin{equation}
\sigma_{\rm el} = \frac {1} {16 \pi \lambda (s, m^2)}
\int^{t^{+}}_{t{-}} {dt \ | T |^2} \ (1-
f_\Phi(\frac{\sqrt{s}}{2}\mp\mu)) \ (1 -
f_\Phi(\frac{\sqrt{s}}{2}\mp \mu))~, \label{qq_cross_section}
\end{equation}
where the inserted blocking factors $(1-f_{\Phi})$ take into
account that rescattered particles appear in a medium where other
identical particles have already existed. Here
$f_\Phi(\sqrt{s}/2\mp\mu) \to f^\mp_{\Phi}$ is the modified Fermi
function (\ref{fermi}) and the sign in front of the chemical
potential $\mu$ corresponds to a particle or an antiparticle. The
integration limits in Eq.~(\ref{qq_cross_section}) for the $qq$
case of equal masses are $ t^+ = 0 $ and $ t^- = 4m^2 - s $. The
kinematic boundary reads $ s> 4m^2$.

For calculation of the differential cross section in the
center-of-mass system it  can be rewritten in the Mandelstam
variables as $t = -2 {p^*}^2 (1-{\rm cos}\Theta)$ and $u = -2
{p^*}^2 (1 + {\rm cos}\Theta) $, where in the center-of-mass
system we have
\begin{equation}
p^* = q^*_1 = q^*_2 = q^*_3 = q^*_4 = \frac{\lambda^{1/2}(s,
m^2, m^2)}{2\sqrt{s}}.
\label{cms}
\end{equation}

It is easy to show that the scattering amplitude for the process
$q \overline {q} \rightarrow q \overline {q}$ can be obtained from
the expressions for the process $qq \rightarrow qq$ by the
substitution $ t \leftrightarrow t $, $ s \leftrightarrow u $ , $
u \leftrightarrow s $. The total elastic scattering amplitude for
$ \overline{q}\ \overline{q} \rightarrow \overline{q}\
\overline{q} $   follows immediately from the $qq\to qq$ amplitude
by time reversal invariance, as long as no chemical potential is
involved.

\subsection*{Quark-antiquark elastic scattering}
\begin{figure} [h]
\centerline{
\includegraphics[width = 4.cm] {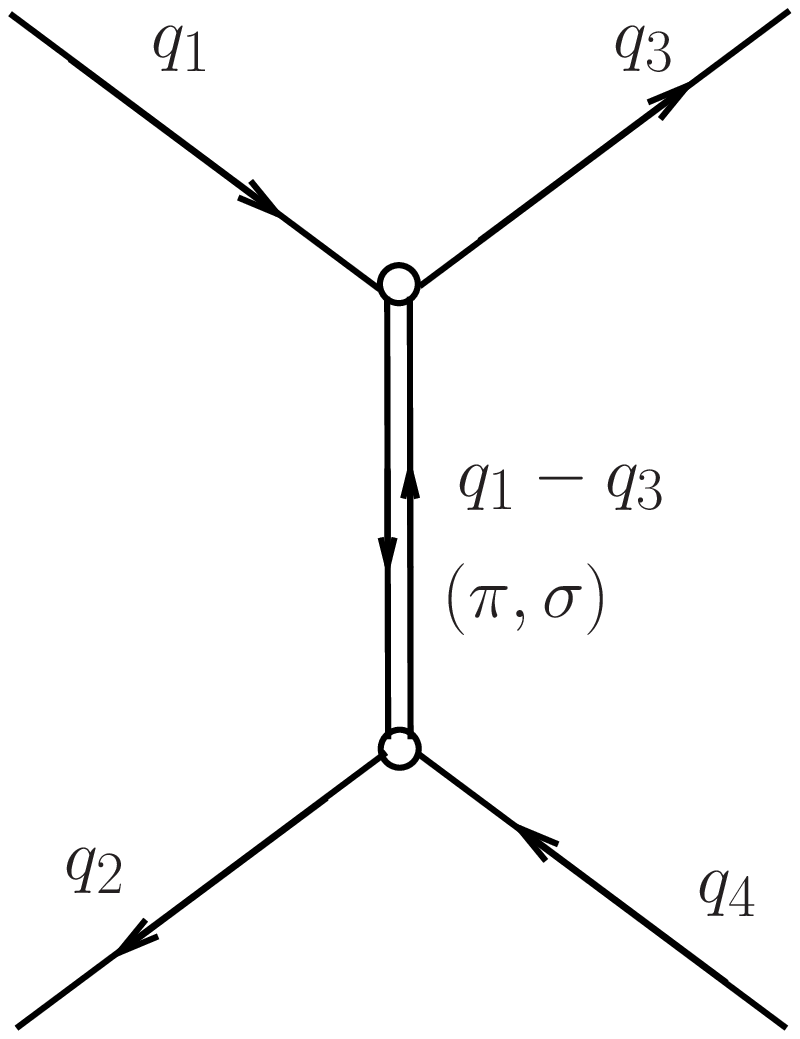}
\hspace{1cm}
\includegraphics[width = 4.cm] {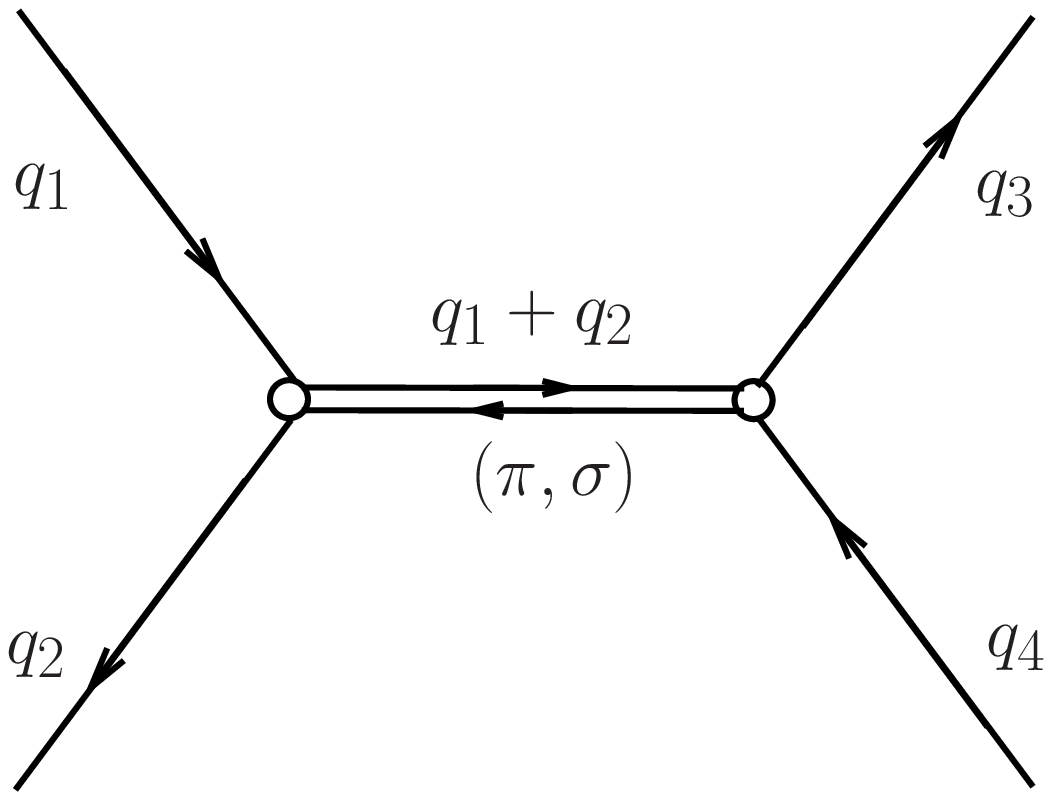}
} \caption{Diagrams of $q\bar{q} \rightarrow q\bar{q} $ for
$t$-channel (left) and $s$-channel (right)} \label{fig_diagqaq}
\end{figure}

 Diagrams for the quark-antiquark process,
$q \overline {q} \rightarrow q \overline {q}$, are given in
Fig.~\ref{fig_diagqaq}.

Similarly to the quark-quark case we have for the $q\bar{q}$
scattering
\begin{eqnarray}
|T_{t}|^2 &=&\left( |D_t^\sigma |^2(t-4m^2)^2 + |D_t^\pi |^2 t^2 \right),\\
|T_{s}|^2 &=& \left( |D_s^\sigma |^2(s-4m^2)^2 + |D_s^\pi |^2 s^2 \right),\\
T_t T^*_s &=& -\frac{1}{2 N_c}( D_t^\sigma D_s^\sigma (ts + 4
m^2(s+t) - 16m^2)
- D_t^\sigma D_s^\pi s (t- 4m^2) \nonumber \\
& - & D_t^\pi D_s^\sigma t (s - 4m^2) + D_t^\pi D_s^\pi ts )
\end{eqnarray}
with the meson propagators
\begin{equation}
D^{\sigma/ \pi}_t = \frac{g^2_{\sigma qq/ \pi qq}}{t - M_{\sigma/ \pi}^2}, \ \  \
D^{\sigma/ \pi}_s = \frac{g^2_{\sigma qq/ \pi qq}}{s - M_{\sigma/ \pi}^2},
\end{equation}
where we   use the pole approximation for meson propagators.

Taking into consideration the isospin factors we can consider the
following independent types of scattering reactions for
quark-quark scattering :
\begin{eqnarray} \label{qq}
&& uu\rightarrow uu \ \ (dd \rightarrow dd), \nonumber \\[-0.1cm]
\\[-0.25cm]
&& ud\rightarrow ud \ \ (du \rightarrow du), \nonumber
\end{eqnarray}
where the first of them includes both $t$ and $u$ channels and the
second includes only the $u$-channel. In the same way, for
quark-antiquark scattering we have a similar relation:
\begin{eqnarray} \label{qq-qq}
&& u\bar{u}\rightarrow u\bar{u} \ \ (d\bar{d} \rightarrow d\bar{d}), \nonumber \\
&& u\bar{d}\rightarrow u\bar{d} \ \ (d\bar{u} \rightarrow d\bar{u}),  \\
&& u\bar{u}\rightarrow d\bar{d} \nonumber
\end{eqnarray}
 with both $t$ and $s$ channels for the first reaction, the $t$-channel for
the second one and the $s$-channel for the third one,
respectively.

The total and differential cross sections for antiquark-antiquark
coincide with those for quark-quark scattering and can be
calculated as in Eqs.~(\ref{dif_cs}), (\ref{qq_cross_section}).

\section{Amplitudes for the $qH \longrightarrow qH$ process }

\begin{figure} [h]
\centerline{\vspace*{1.cm}
\includegraphics[width=4.5cm] {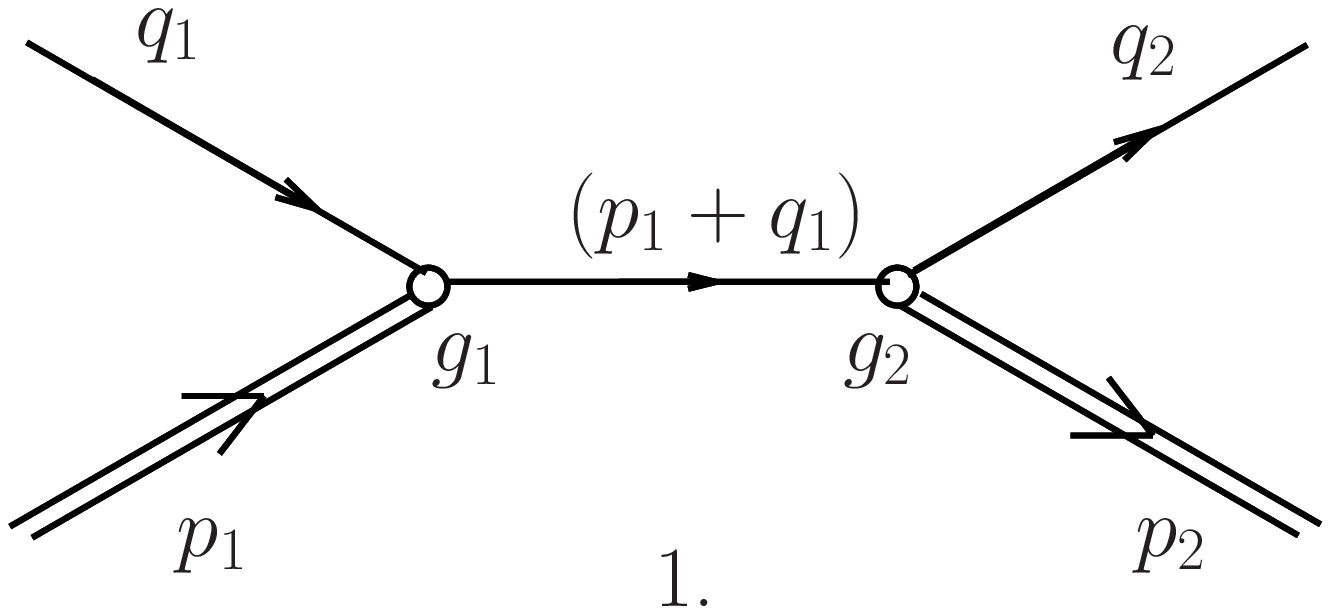}
\hspace{0.5cm}\vspace*{-1.cm}
\includegraphics[width=3.5cm]{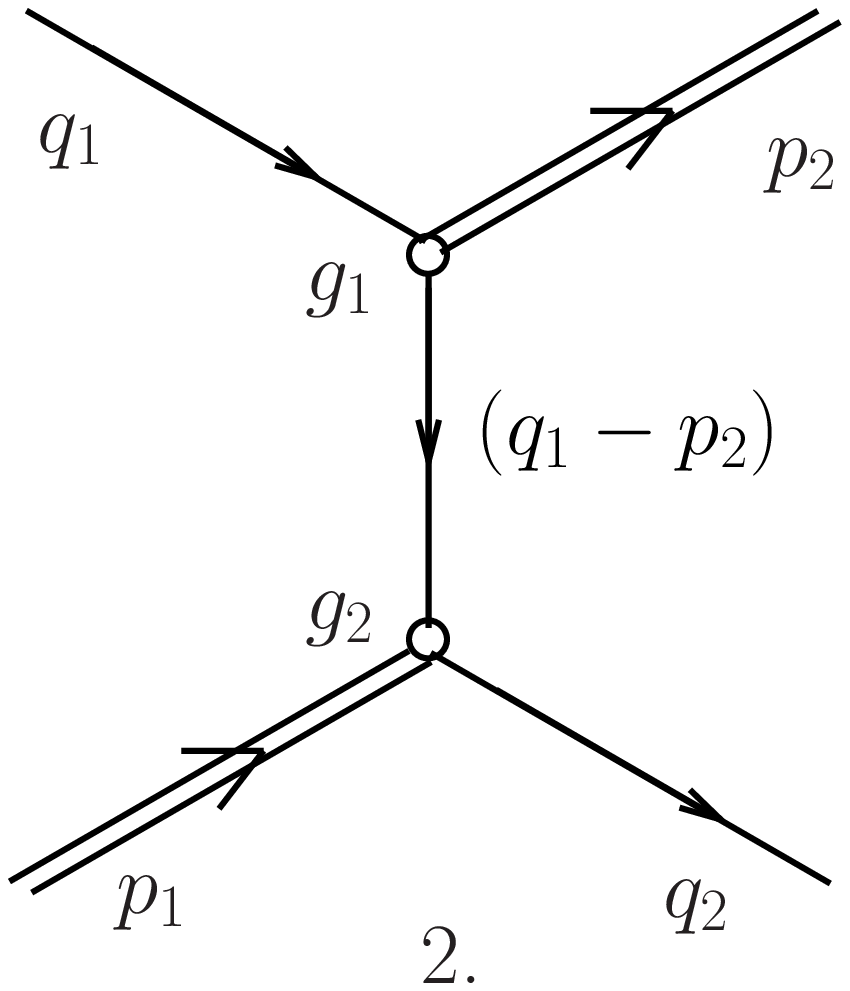}
\hspace{0.5cm}
\includegraphics[width=3.5cm]{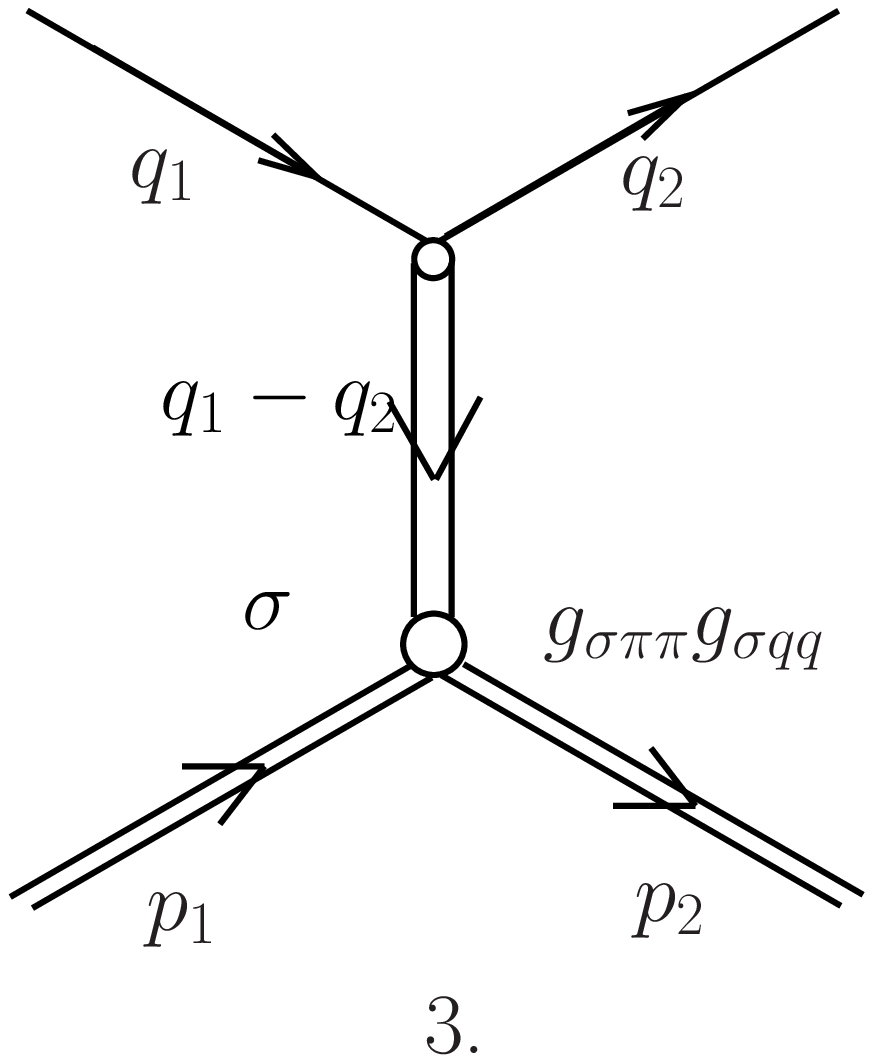}}
\caption{The Feynman diagrams for the $s-$ and $u-$quark exchange
channels (diagrams 1 and 2) and the t-channel with the exchange
 by the sigma-meson (diagram 3). The double line corresponds to a
pion.} \label{fig_Dir}
\end{figure}

The Feynman diagrams for calculation of the quark-pion scattering
amplitudes with the exchange of a quark, $\pi$- and $\sigma
$-meson are shown in Fig.~\ref{fig_Dir}.

The amplitude in the $s$-channel (diagram 1 in Fig. \ref{fig_Dir})
is as follows:
\begin{eqnarray}\label{T1}
-i T_1 &=& \overline {u} ({\bf q_2}) g_2 (i \gamma_5)
\frac{((\widehat {q} _1 + \widehat {p} _1) + m)} {s-m^2}
(i \gamma_5) g_1 u ({\bf q_1}), \nonumber \\[-0.1cm] \\[-0.25cm]
i T^*_1 &=& u ({\bf q_2}) g_2 (i \gamma_5) \gamma_0 \frac {((\widehat
{q} _1 + \widehat {p} _1) + m)^\dagger} {s-m^2 } \gamma_0 (i \gamma_5)
g_1 \overline {u} ({\bf q_1})~. \nonumber
\end{eqnarray}

After the transformation, Eq.~(\ref{T1}) reads
\begin{eqnarray}
i T^*_1 & = &-iu ({\bf q_2}) g_2 (i \gamma_5) \frac {((\widehat
{q} _1 + \widehat {p} _1) + m)} {s-m^2} (i \gamma_5) g_1 \overline
{u} ({\bf q_1})~.
\end{eqnarray}
The result for the $u$-channel can be obtained by the substitution
$ p_1 \rightarrow-p_2 $
\begin{eqnarray} \label{T2}
- i T_2 & = &  \overline {u} ({\bf q_2}) g_2 (i \gamma_5)
\frac {((\widehat {q} _1-\widehat {p} _2) + m)}
 {u - m ^ 2} (i \gamma_5) g_1 u ({\bf q_1}), \nonumber
 \\[-0.1cm]\\[-0.25cm]
i T ^ * _2 & = & u ({\bf q_2}) g_2 (i \gamma_5) \frac {((\widehat {q}
_1-\widehat {p} _2) + m)} {u - m ^ 2} (i \gamma_5) g_1 \overline {u}
({\bf q_1}).\nonumber
\end{eqnarray}

The amplitude of the process with the $\sigma$-meson  exchange is
as follows~:

\begin{eqnarray}\label{T3}
-i T_3 &=& \overline {u}(q_2) {\bf 1} u(q_2) \frac {1} {t-M ^2_ \sigma} g_ {\sigma \pi \pi} g_ {\sigma qq}, \nonumber
\\[-0.1cm]\\[-0.25cm]
i T^*_3 &=& \overline {u} (q_2) (\gamma_0 {\bf 1} \gamma_0) u (q_2)
\frac {1} {t-M^2_ \sigma} g_ {\sigma \pi \pi } g_ {\sigma
qq}.\nonumber
\end{eqnarray}

The total amplitude of the process is given as $ \displaystyle | T |^2 = f_c \sum_{c} |T_1 +
T_2 + T_3|^2 $ and then after tracing and
transforming equations we get:

\begin{eqnarray}
T_1T^*_1 &=& N_c K_s ^ 2 [M_\pi^4 - (s - m^2)(u - m^2)], \label{11a}\\
T_1T^*_2 &=& T_2T ^ * _1 =N_c K_s K_u [- M_\pi^4 + (s - m^2)(u - m^2)],  \label{12a}\\
T_1T^*_3 &=& T_3T^*_1 = N_c K_s K_t [m (s-u)], \label{13a}\\
T_2T^*_2 &=& N_c K_u ^ 2 [M_\pi^4 - (s - m^2)(u - m^2)], \label{22a}\\
T_3T^*_2 &=& T_2T^*_3 = - N_c K_u K_t [m (s-u)], \label{32a}\\
T_3T^*_3 &=& N_c K_t^2 (4m^2-t)~,\label{33a}
\end{eqnarray}
where $ m, M_\pi, M_\sigma $  are the masses of the quark,  pion
and sigma-meson, respectively. Summation is carried over color and
the type of reacton.

Here  we have introduced the propagators $ K_s, K_u, K_t $
\begin{equation}
K_s = \frac {g^2_{\pi qq}} {s - m ^ 2}, \ \ \
K_u = \frac {g^2_{\pi qq}} {u - m ^ 2},\ \ \
K_t = \frac {g_ {\sigma \pi \pi} g_ {\sigma qq}} {t -M^2_\sigma},
\label{Kt}
\end{equation}
where $g_{\pi qq} $, $ g_ { \sigma qq} $ are defined by Eq.
(\ref{couple}) and the coupling strength of $\sigma -\pi \pi$ is
interrelated  as $ g_{\sigma \pi\pi} = 2 g_{\sigma qq}g^2_{\pi qq}
A_{\sigma \pi \pi}$ and $A_{\sigma \pi \pi}$ is the amplitude of
the $\sigma\rightarrow \pi \pi$ decay~\cite{spipi}. Summation over
colors depends on the reaction type. Taking into account the
diagrams with $1/N_c$ and $1/N_c^2$ corrections we have got color
factors $f_c$. Both color and flavor factors for each reaction are
given in Table \ref{table1}.

\begin{table} [thb]
\caption{\label{table1} Flavor and color structure of the
processes}
\begin{center}
\begin{tabular} { r  c  c }
\hline
Process & Isospin factor & color factor $f_c$\\
\hline
$ \rm{u} \pi^0 \rightarrow \rm{u} \pi^0 $ & $ \frac {1} {2} K_s $,\ $ \frac {1} {2} K_u $ ,\ $ K_t $\ & $\left(1 + \frac{2}{N_c} + \frac{1}{N_c^2}\right)$ \\
$ \rightarrow \rm{d} \pi^+ $ &  $ \frac {1} {\sqrt {2}} K_s $,\ $ \frac {1} {\sqrt {2}} K_u $,\ $ K_t $ & $\left(1 + \frac{1}{N_c}\right)$ \\

$ \rm{u} \pi^- \rightarrow \rm{u} \pi^- $ &  $ K_s $,\ $ K_u $,\ $ K_t $ & $\left(1 + \frac{1}{N^2_c}\right)$\\
$ \rightarrow \rm{d} \pi ^ 0 $ & $ \frac {1} {\sqrt {2}} K_s $,\ $ \frac {1} {\sqrt {2}} K_u $,\ $ K_t $\ & $\left(1 + \frac{1}{N_c}\right)$ \\

$ \rm{d} \pi^0 \longrightarrow \rm{d} \pi^0 $ &  $ \frac {1} {2} K_s $,\ $ \frac {1} {2} K_u $,\ $ K_t $\ & $\left(1 + \frac{2}{N_c} + \frac{1}{N_c^2}\right)$\\
$ \rightarrow \rm{u} \pi^- $ &  $ \frac {1} {\sqrt {2}} K_s $, \ $ \frac {1} {\sqrt {2}} K_u $,\ $ K_t $ \ & $\left(1 + \frac{1}{N_c}\right)$\\

$ \rm{d} \pi^+ \longrightarrow \rm{d} \pi^+ $ &  $ K_s $,\ $ K_u $,\ $ K_t $\ & $\left(1 + \frac{1}{N^2_c}\right)$\\
$ \rightarrow \rm{u} \pi^0 $ &  $ \frac {1} {\sqrt {2}} K_s $,\ $ \frac {1} {\sqrt {2}} K_u $,\ $ K_t $\ & $\left(1 + \frac{1}{N_c}\right)$ \\
\hline
\end{tabular} \label{table1}
\end{center}
\end{table}

Kinematic invariants for the $qH \rightarrow qH$ scattering
process are defined as follows:
\begin{eqnarray}
&& s = (q_1 + p_1) ^ 2 = (q_2 + p_2)^2, \nonumber \\
&& t = (q_1 - q_2)^2 = (p_1 - p_2)^2, \nonumber
\\[-0.1cm]\\[-0.25cm]
&& u = (q_1 - p_2)^2 = (q_2 - p_1)^2, \nonumber \\
&& s + t + u = 2 M_\pi^2 +  2 m^2. \nonumber
\end{eqnarray}

The differential cross section has the form similar to
Eq.~(\ref{dif_cs})
\begin{equation}
\frac{d\sigma_{el}}{dt} = \frac {| T |^2} {16 \pi \lambda (s, m^2,
M_\pi^2)} \label{dif_cs1}
\end{equation}
with $ \lambda(s, m^2, M_\pi^2) = (s - (M_\pi + m)^2) (s - (M_\pi
- m)^2)$. Accordingly, the  integrated cross section for the
elastic $qH$ scattering is:
\begin{equation}
\sigma_{\rm el} = \frac {1} {16 \pi \lambda (s, m^2, M_\pi^2)}
\int^{t^{+}}_{t{-}} {dt \ | T |^2}(1 - f_\Phi(E_q\mp\mu)) \ (1 +
f_{B}(E_H\mp \mu))~, \label{cross_section1}
\end{equation}
where $E_q$, $E_H$ correspond to the quark and hadron energies,
the Bose-Einstein factor has the form $f_B = (\exp(\beta x)
-1)^{-1}$ and  the integration limits are
\begin{equation}
t^\pm = 2 m^2 - \frac{1}{2s}\left\lbrace (s+m^2 -
M_\pi^2)^2 \mp \lambda(s, m^2, M_\pi^2) \right\rbrace.
\end{equation}
As follows from Eq.~(\ref{cross_section1}) for the cross section,
the $q\pi$ reaction has kinematic boundaries $s>{\rm
max}\left\lbrace (m+M_\pi)^2, (m-M_\pi)^2\right\rbrace $.

For calculation of the differential cross section in the
center-of-mass system we can use the same expressions as for the
$qq$-scattering keeping in mind that ${p^*}^2$  differs from
Eq.~(\ref{cms}) because the $\lambda(s, m^2, M_\pi^2)$ factor has
a different form.

\section{Numerical results and discussion}

\subsection*{Quark-(anti)quark elastic cross-section}
\begin{figure} [!h]
\centerline{
\includegraphics[width = 7.2cm] {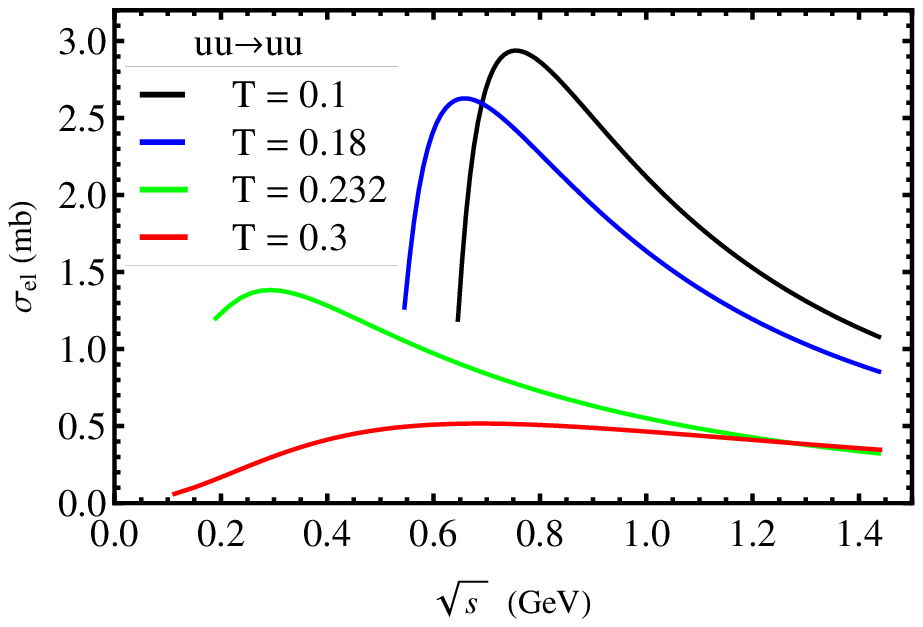}
\includegraphics[width = 7.2cm] {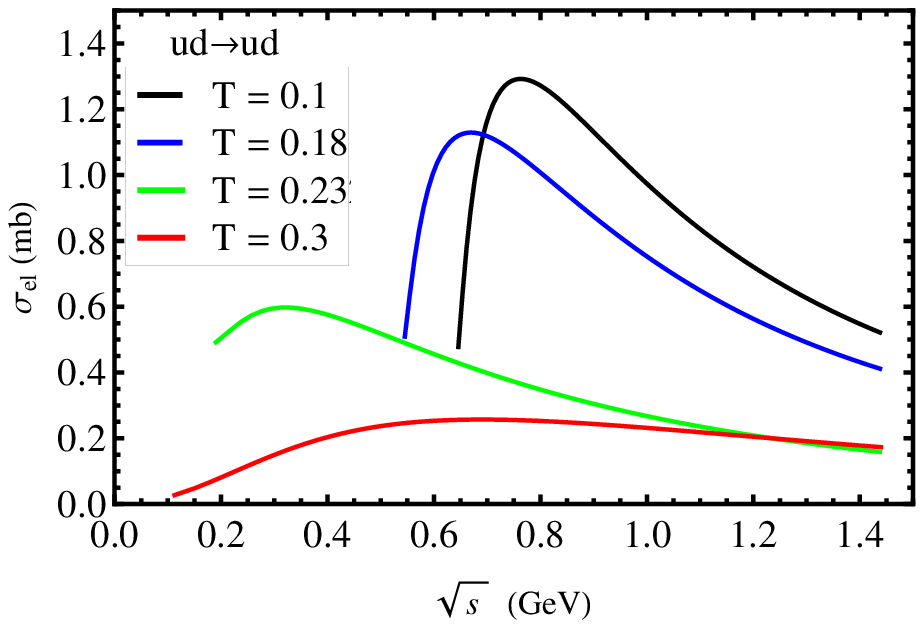}}
\centerline{
\includegraphics[width = 7.2cm] {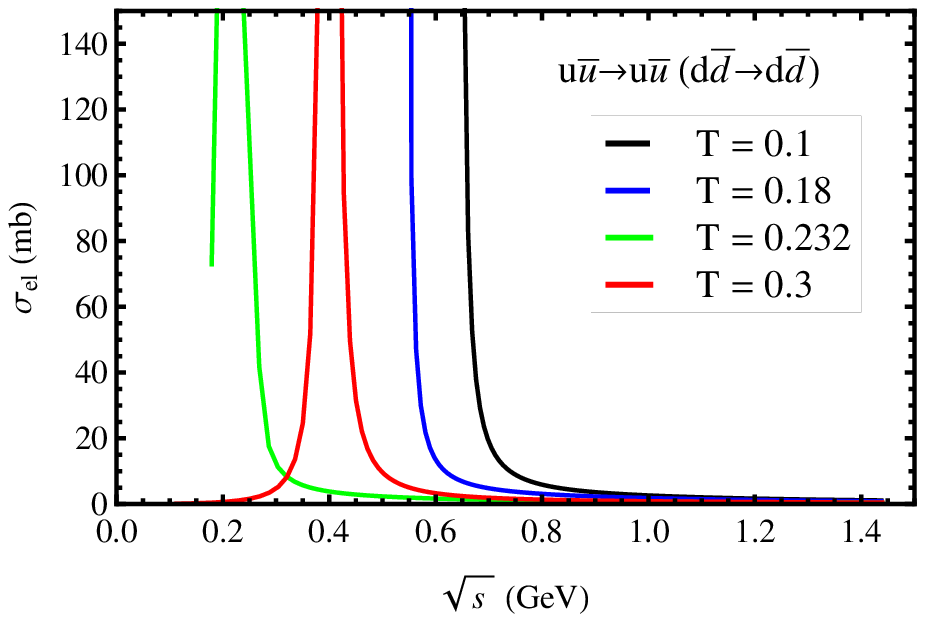}
\includegraphics[width = 7.2cm] {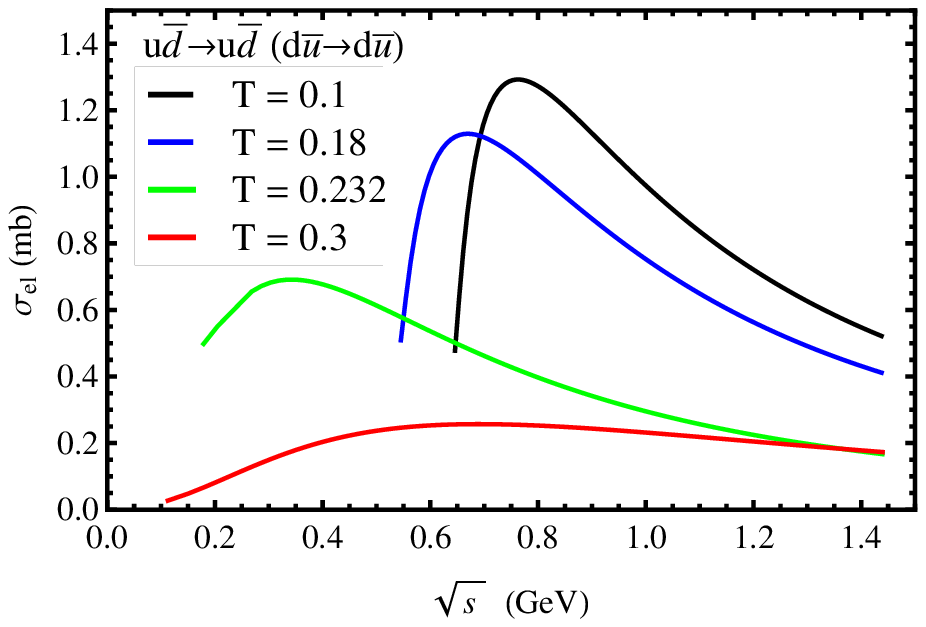}}
\caption{Total cross section of reactions $qq \rightarrow qq$ (two
upper plots) and $q\bar{q}\rightarrow q\bar{q}$ (two bottom plots)
at different temperatures.} \label{crossqq}
\end{figure}

We shall start with the PNJL calculation of the in-medium $qq,
q\bar{q}$ and $\bar{q}\bar{q}$ cross sections.  The integral cross
sections $\sigma_{el}$ are plotted in Fig.~\ref{crossqq} for
various types of reactions shown in Eqs.~(\ref{qq}),(\ref{qq-qq}).
A strict limit of the applicability of the quark model at high energies
is not well established. As mentioned in work \cite{RKH96}, the center-of-mass energy $\sqrt{s}$ in our model is restricted by the scale $\sqrt s= 2 \sqrt {\Lambda^2+m_q^2} \approx$ 1.5 GeV.

 The quark scattering cross sections are relatively
featureless. For quark-quark scattering  the flavor influences
mainly the magnitude of the cross sections, being larger for
quarks of the same flavors ($\sigma_{el}(uu\to uu)\approx 2.5
\sigma_{el}(ud\to ud)$) (compare two upper panels in
Fig.~\ref{crossqq}). This is because the $ud$ scattering has  in
fact less exchange mesons available in the $u$-channel, since no
neutral particles are admissible in this channel. The energy
dependence of $\sigma_{el}$ is very similar in both cases: the
cross section has a clear maximum at the energy $\sqrt{s}\sim$1
GeV for a moderate temperature $T\sim$ 0.1 GeV which moves to
smaller $\sqrt{s}$ as temperature increases exhibiting almost
singular behavior at $T=T_{Mott}$.  As noted in Introduction,
the PNJL  model is reliable for temperatures $T\lsim 2.5
T_c$~\cite{MOM04} where for our 2-flavor model we have $T_c\approx
0.5 (T_0+T_{Mott}) =$210 MeV (see Tabl~\ref{param}). The
$\sigma_{el}$ is getting rather flat if the temperature exceeds
the Mott temperature (see the case $T=$ 0.3 GeV in
Fig.~\ref{crossqq}).

 Note that the cross section is evaluated according to
Eq.~(\ref{qq_cross_section},) where the modified Pauli factor for
scattered quarks is involved. As is seen from Fig.~\ref{crossqq},
$\sigma_{el}$ decreases when $\sqrt{s}$ grows, which is a
consequence of nonperturbativity in the coupling constant $G$ of
the PNJL treatment. In the Born approximation the opposite
behavior $\sigma_{el}\sim s$ is observed, as was demonstrated in
Ref.~\cite{ZHKN95}. The behavior of the scattering cross section
for quark-antiquark of different types is very close to that for
quark-quark of different flavors (compare $ud$ and $u\bar{d}$
reactions in Fig.~\ref{crossqq}). However, for quark-antiquark of
the same flavor ($u\bar{u}$ and $d\bar{d}$) the cross section at
any temperature has the resonance-dominated behavior demonstrating
a huge maximum located at the energy close to the $\sigma$-meson
mass. In other words, the quark-antiquark scattering shows a
threshold divergence at the Mott temperature $T_{Mott}$, at which
the pion dissociates into its constituents and  becomes a resonant
state. This feature manifests itself in other processes like
$q\bar{q}\to \gamma\gamma$~\cite{RKB97} as well as
$\pi\pi\to\pi\pi$~\cite{spipi,Quack95} and
$\pi\gamma\to\pi\gamma$~\cite{DVHKR97}. The dramatically high
cross sections mean that near $T_{Mott}$ a local equilibrium can
be established.

\begin{figure} [h]
\centerline{
\includegraphics[width = 8.cm] {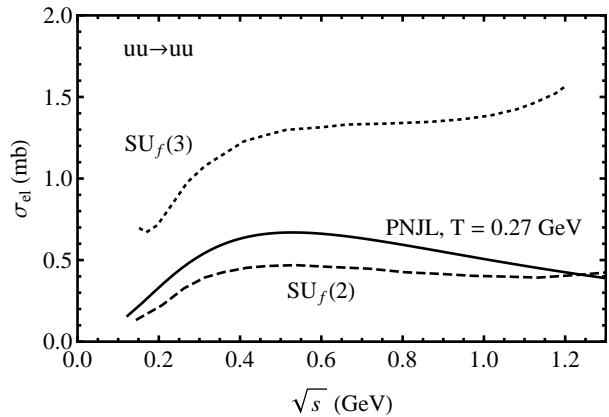}}
\caption{Comparison of the energy dependence of the elastic cross
section for the $uu \rightarrow uu$ scattering for a temperature
above the Mott temperature. The dotted and dashed lines are the
NJL results for $SU_f(3)$ and $SU_f(2)$, respectively, from
Refs.~\cite{RKH96PC,ZHKN95}, the solid line is our two-flavor PNJL
calculations for $T=$ 0.27 GeV.}
 \label{uu-comp}
\end{figure}
The calculated results can be compared with those obtained
earlier~\cite{ZHKN95,RKH96}. Some elastic scattering cross
sections were calculated in the two-flavor sector $SU_f(2)$ of the
NJL model with the additional restriction of the chiral limit
condition $m =$0 in Ref.~\cite{ZHKN95}. The three flavor NJL model
was considered in~\cite{RKH96} and the results are exemplified in
Fig.~\ref{uu-comp}. It is seen that the $SU_f(3)$ calculation
yields a larger cross section for $uu \rightarrow uu$ than the
corresponding two-flavor case. This is due to the point  that the
additional exchange channels $\eta, \eta'$ and $\sigma'$ are
missing in the two flavor case. In particular, the $uu \to uu$
channel of the elastic scattering at $T = 0.215$ GeV for $SU_f(3)$
is regularly above that for $SU_f(2)$ by the factor of (3-4) in
the whole energy range from the threshold till $\sqrt{s}\sim$1.2
GeV~\cite{RKH96} (compare two dotted lines in Fig.~\ref{uu-comp}).
The values of the $uu \to uu$ cross section for
$SU_f(2)$~\cite{RKH96} with $T_{Mott}$ are consistent with our
PNJL results presented in Fig.~\ref{uu-comp} if one takes into
account the different Mott temperatures in these calculations. It
is not surprising since the suppression due to the Polyakov loop
works only in the low temperature sector, $T<T_{Mott}$.

\begin{figure} [h]
\centerline{
\includegraphics[width = 7.2cm] {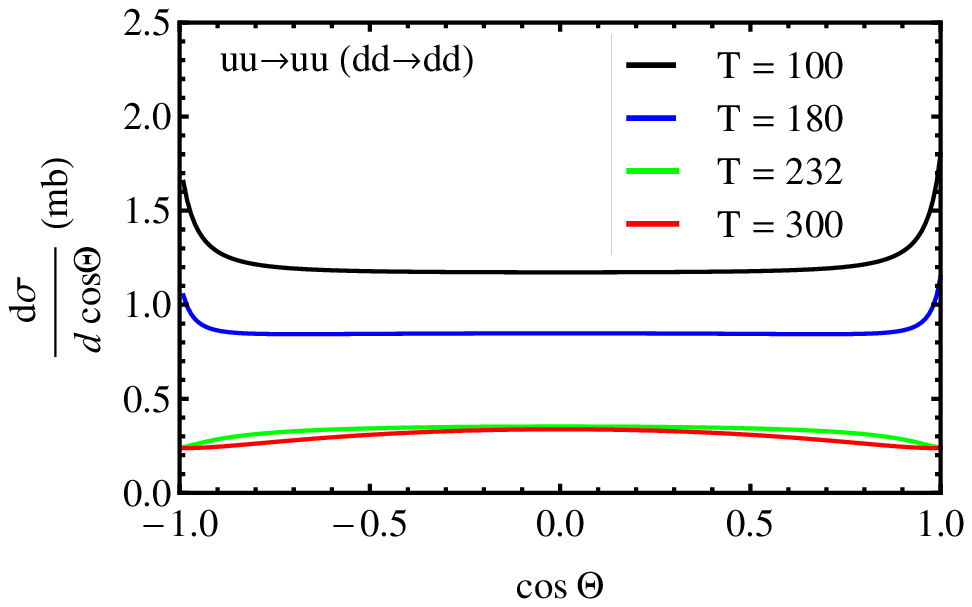}
\includegraphics[width = 7.2cm] {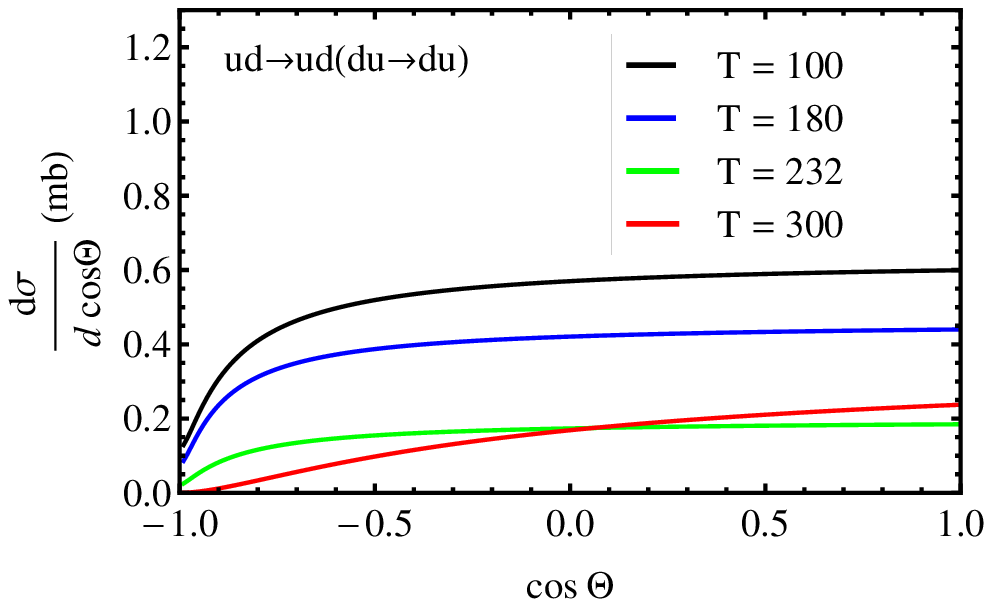}}
\centerline{
\includegraphics[width = 7.2cm] {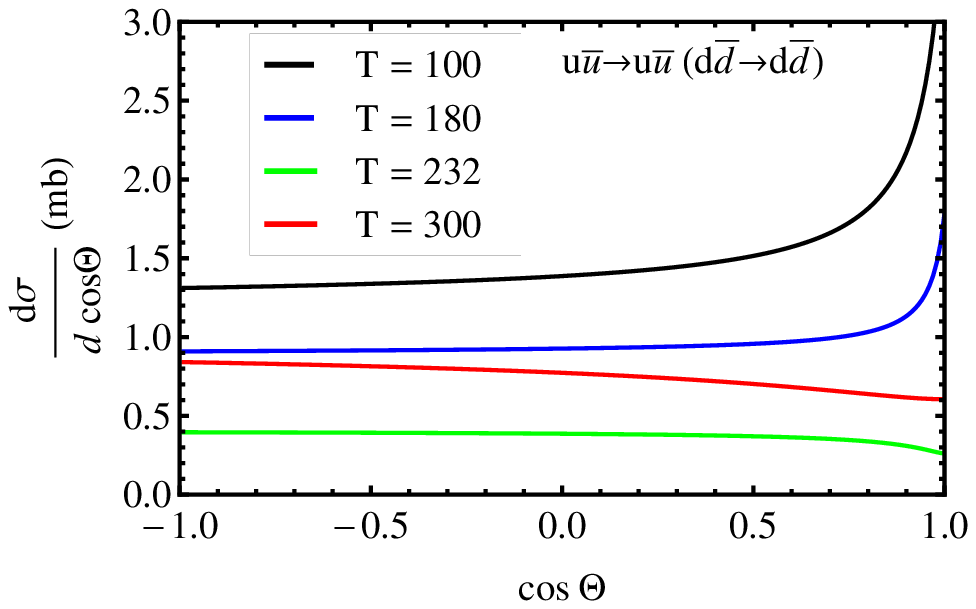}
\includegraphics[width = 7.2cm] {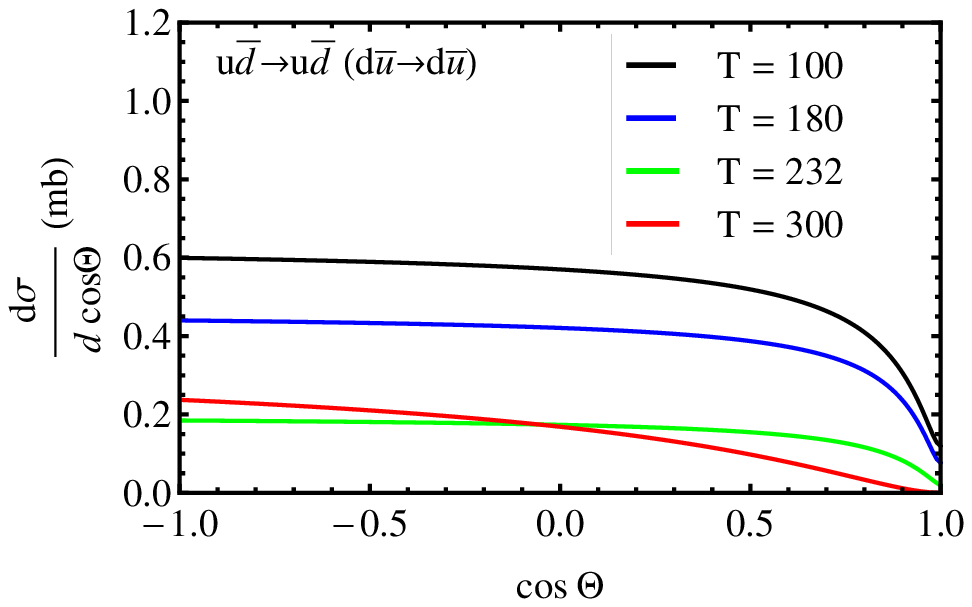}}
\caption{Angular distribution of the elastic scattering processes
$qq \rightarrow qq$ (upper panels) and $q\bar{q}\rightarrow
q\bar{q}$ (bottom panels) at $\sqrt{s} = 1$ GeV for different
temperatures. } \label{adqq}
\end{figure}

 For each process the differential cross sections may also be calculated.
As an example, the angular distribution in the elastic $qq$ and
$q\bar{q}$ scatterings is presented in Fig.~\ref{adqq} at the
energy $\sqrt{s}=$1 GeV. In the case of quarks of the same flavors
($uu$), the angular distribution is isotropic except for the
narrow region $|cos \Theta| \gsim $0.9. The differential
distributions for quarks of various flavors ($ud$) or different
types ($u\bar{d})$ are also almost flat, besides the region $|cos
\Theta| \gsim $0.8, but this deflection from isotropy is in the
backward direction for the $ud$ while it is in the forward
direction for $u\bar{d}$ reactions. For the $u\bar{u}$ and the
$d\bar{d}$ scatterings, the angular distribution changes from
isotropic to clearly anisotropic one if the temperature decreases
from $T_{Mott}$ to about 0.1 GeV. In accordance with the
temperature dependence of $\sigma_{el}$, the magnitude of the
differential distribution of $d\sigma_{el}/dcos \Theta $ decreases
with the growth of the temperature.

The processes presented in Figs.~\ref{fig_diagqq} and
\ref{fig_diagqaq} can also be calculated in the lowest order
perturbative QCD. These on-shell high-energy calculations are in
use starting  from the first classical parton cascade
model~\cite{Ge95}.  In Fig.~\ref{pQCD}, the comparison between the
PNJL and the pQCD results (obtained according to the model
description in Appendix of Ref.~\cite{ZHKN95}) is given. As is
seen, in contrast with the PNJL, the pQCD description predicts a
strong scattering enhancement at forward angles  for finite parton
masses. In the limit $m\to$0 the cross section increases and
diverges.
\begin{figure} [h]
\centerline{
\includegraphics[width = 8cm] {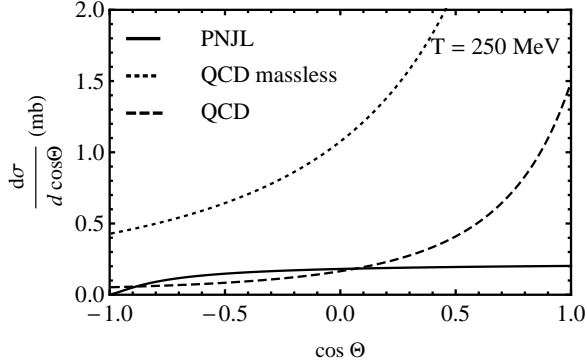}}
\caption{Comparison of the differential $ud\to ud$ cross section
calculated in the pQCD model at $\sqrt{s}=$1 GeV for finite
(dashed line) and vanishing (dotted line) masses. Our results for
$T=$ 0.25 GeV are plotted by the solid line. }
 \label{pQCD}
\end{figure}

Such behavior is more transparent from  simplified two-body pQCD
calculations with $m=0$ used in the AMPT model~\cite{AMPT}. In
particular, for the gluon elastic scattering (which differs by the
Casimir factor from the $qq$ scattering) one has
\begin{equation}\label{el_QCD}
\frac{d\sigma_{gg}}{dt}=\frac{ 9\pi \alpha^2_s}{ 2
s^2}(3-\frac{ut}{s^2}-\frac{us}{t^2}-\frac{st}{u^2})
 \simeq \frac{ 9\pi \alpha^2_s}{2} (\frac{1}{t^2}+\frac{1}{u^2})  \simeq
\frac{ 9\pi \alpha^2_s}{ 2 t^2} ,
\end{equation}
with the strong coupling constant $\alpha_s$.
Equation~(\ref{el_QCD}) is obtained in the leading-order QCD by
keeping only the leading divergent terms for identical particles,
which allows one to limit oneself to the angle range
$0\leq\Theta\leq\pi/2$ (the last equality in (\ref{el_QCD})). The
cross section really diverges at the scattering angle $\Theta=$0.
The singularity in this cross section can be regularized by
introducing the Debye screening mass $M_D$ leading to
\begin{eqnarray} \label{d-pQCD}
\frac{d\sigma_{gg}}{dt}= \frac{ 9\pi \alpha^2_s} { 2(t-M_D^2)^2}
\end{eqnarray}
and, respectively, for the total cross section at relativistic
energies
\begin{eqnarray}\label{s-pQCD}
\sigma_{gg}\simeq \frac{ 9\pi \alpha^2_s}{ 2M_D^2}
\frac{1}{1+M_D^2/s}\approx \frac{ 9\pi \alpha^2_s}{ 2 M_D^2}~,
\end{eqnarray}
if $s\gg M^2_D$~\cite{AMPT}. Note that in this approximation the
pQCD gives an energy-independent cross section (\ref{s-pQCD}) in
disagreement with the NJL-like chiral model discussed. Taking
$M_D=$3 fm$^{-1}$ we get $\sigma_{el}=$3 mb. In the real AMPT
calculations $M_D$ is a parameter and this cross section  changes
from 3 to 10 mb in different model versions~\cite{AMPT}.

\subsection*{Quark-hadron elastic cross-section}

\begin{figure} [thb]
\centerline{
\includegraphics[width = 7.2cm] {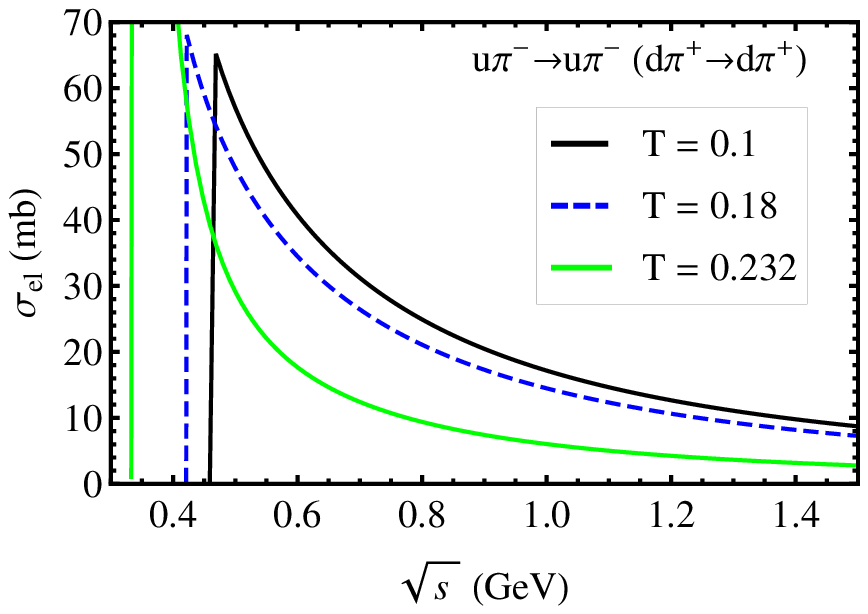}
\includegraphics[width = 7.2cm] {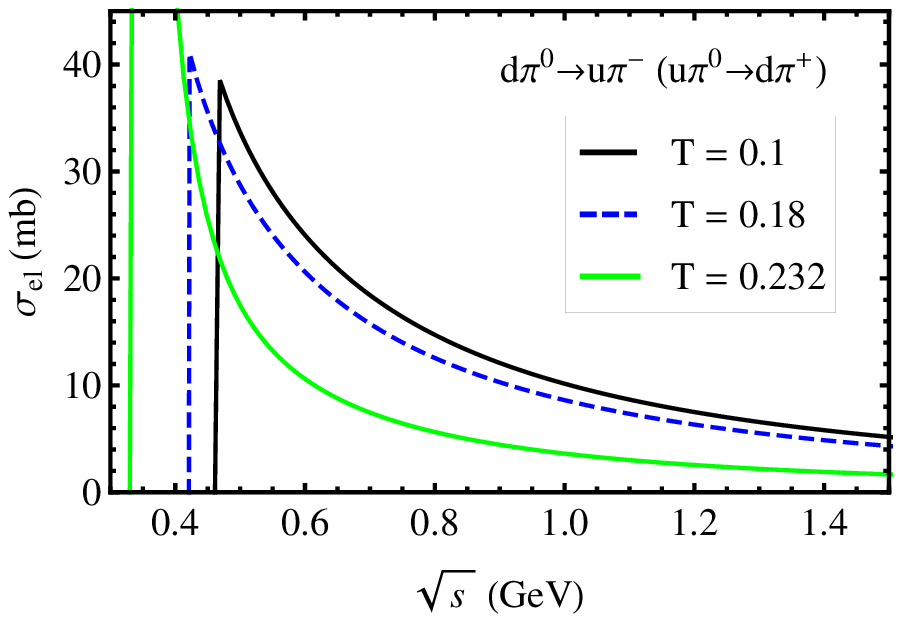}}
\centerline{ \hspace*{-5mm}
\includegraphics[width = 7.2cm] {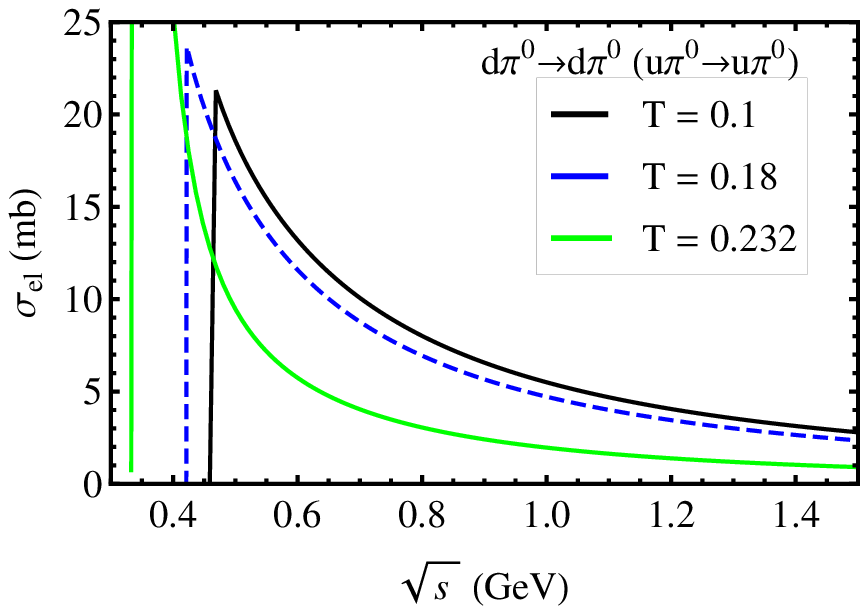}
\includegraphics[width = 6.2cm] {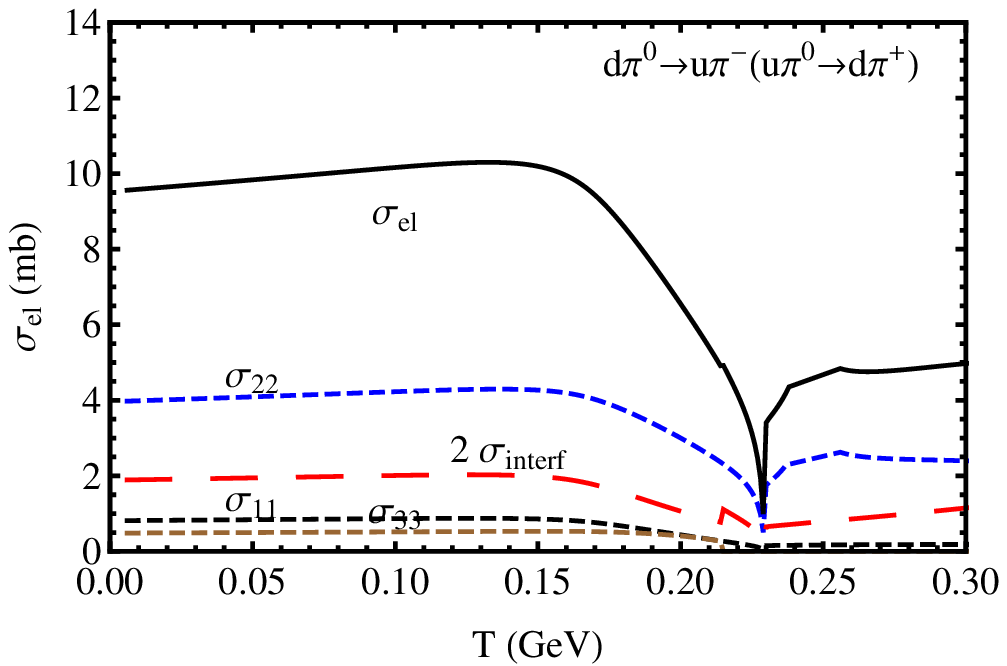}}
 \caption{Energy dependence of the total scattering cross
sections for the processes $ qH \rightarrow qH $ corresponding to
the different reactions and the temperature dependence of
different components (see the text) of the $d\pi^0\to u\pi^-$
scattering cross section at $\sqrt{s}=$1 GeV (right-bottom panel).
}
 \label{crosqh}
\end{figure}

The integral cross section for the in-medium elastic scattering of
quarks on pions is presented in Fig.~\ref{crosqh}. For the
considered reactions $u\pi^-, d\pi^+, u\pi^0$ and their charge
conjugated ones the cross sections behave very similarly:
$\sigma_{el}$ is maximal just at the threshold
$s_{thr}=(m+m_\pi(T))^2$ reaching here values of 60-80 mb  and
then monotonically falls down rather quickly with the slightly
different slopes  for different reactions. At the Mott temperature
and for all energies the $qH$ elastic cross section exhibits a
huge maximum as large as several hundred of mb (to be cut in
Fig.~~\ref{crosqh}). Generally, the magnitude of the quark-pion
cross sections is higher than the quark-quark one and comparable
with free hadron-hadron cross sections. In the case of the
$u\pi^-$ reaction, the cross section fall-off is the slowest, and
the quark-pion scattering turns out to be several times higher
than that for quark-quark in the energy range $\approx$(0.7-1.5)
GeV.

As is seen from Fig.~\ref{crosqh} (right-bottom panel),
$\sigma_{el}(T)$ is a weakly changing function up to $T\sim$ 0.18
GeV and then has a kick at the Mott temperature. This cross
section is the sum of different terms corresponding to the squared
amplitudes: $\sigma_{11}$ (see Eq.~(\ref{11a})), $\sigma_{22}$
(Eq.~(\ref{22a})), $\sigma_{33}$ (Eq.~(\ref{33a})) and summary
interference term $\sigma_{interf}$
(Eqs.~(\ref{12a})+(\ref{13a})+(\ref{32a})). The dominant
$u$-channel gives $\sigma_{22}\sim 0.5\ \sigma_{el}$. In contrast,
the cross section of the $t$-channel is very small,
$\sigma_{33}\simeq$0.5 mb. The rest of $\sigma_{el}$ is shared
almost equally between the sum of all interference terms
$\sigma_{interf}$ (note the scale in figure) and the $s-$channel
$\sigma_{11}$ elastic cross section.

\begin{figure} [thb]
\centerline{
\includegraphics[width = 7.2cm] {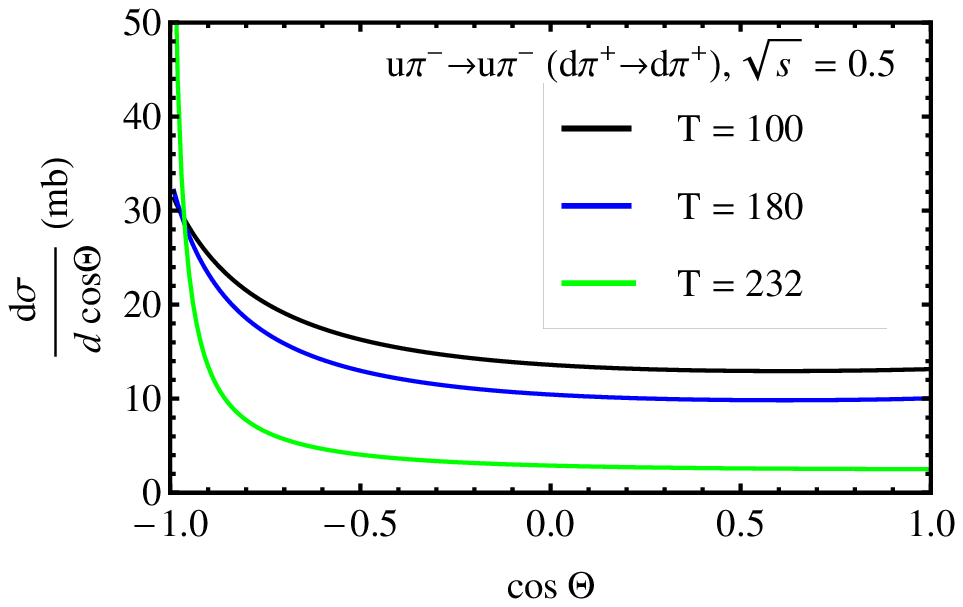}
\includegraphics[width = 7.2cm] {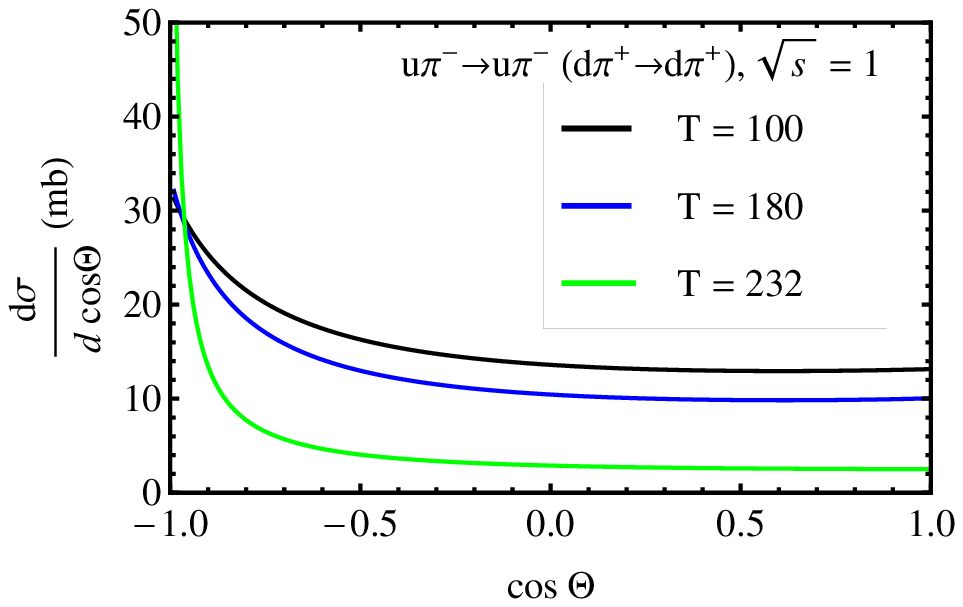}}
\centerline{
\includegraphics[width = 7.2cm] {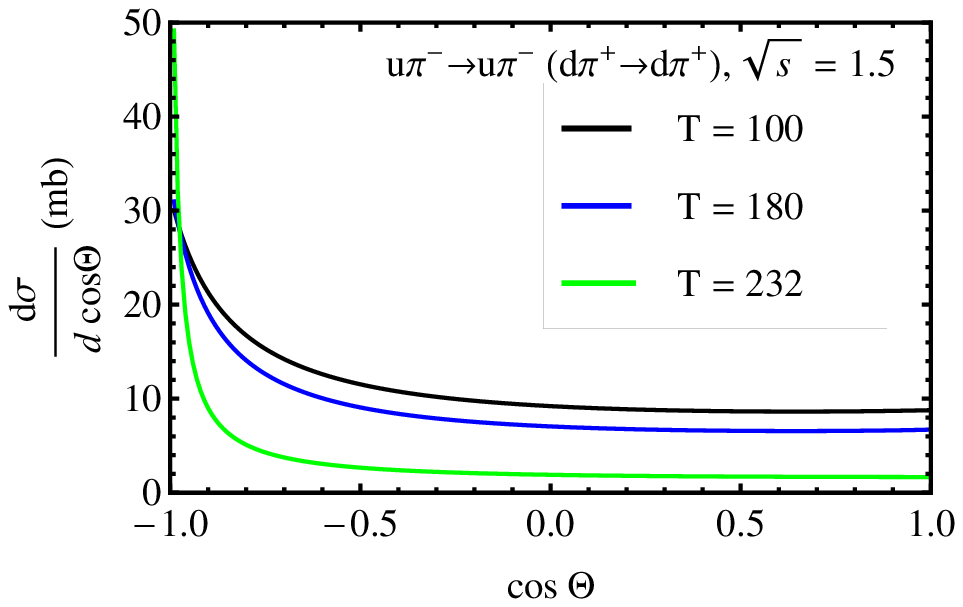}
\includegraphics[width = 6.8cm] {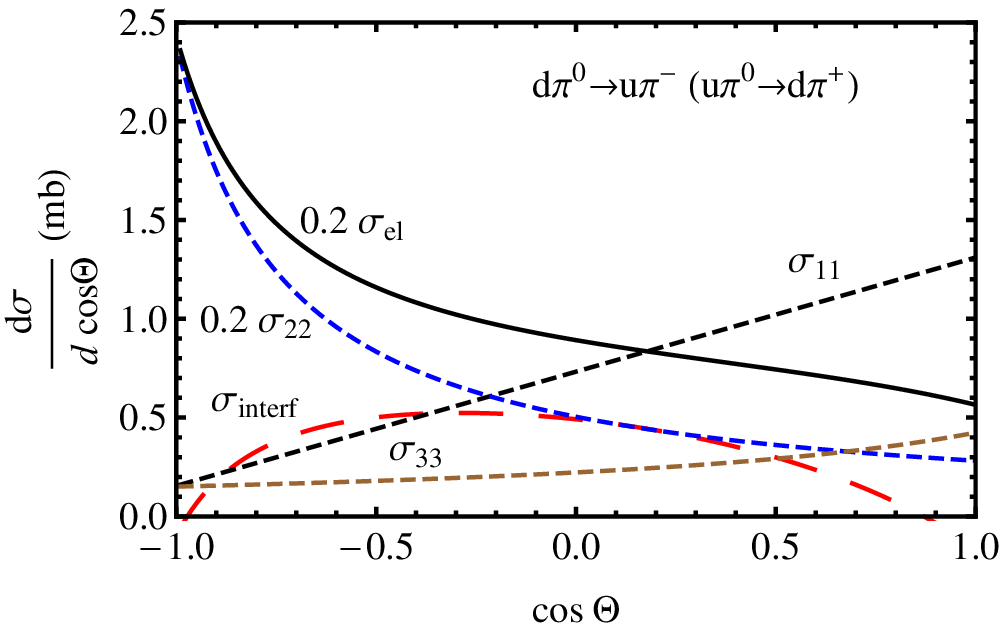}}
\caption{Angular distributions of the $u\pi^- \rightarrow u\pi^-$
scattering for all reactions at various temperatures and energies
and the contributions of different channels of the $d\pi^0\to
u\pi^-$ reaction  to $d\sigma_{el}/d cos\Theta$ for $\sqrt{s}=$1
GeV and $T=$ 0.1 GeV (right-bottom panel). Notation of components
is the same as in Fig.~\ref{crosqh}. }
    \label{adqh}
\end{figure}

In Fig.~\ref{adqh}, some examples of the angular distribution are
given for the quark-pion scattering. At the energy $\sqrt{s}\lsim$
0.5 GeV, independently of T, the angular distributions are
isotropic with high accuracy, besides in the vicinity of
$T_{Mott}$. At higher energy and moderate temperatures
$T<T_{Mott}$ the differential distributions are almost isotropic
with small anisotropy growing at the backward scattering angles.
This backward bounce is more prominent for lower energies. The
presented angular distributions are really the contributions of
diagrams of three types (see Fig.~\ref{fig_Dir}). As is shown in
the right-bottom panel of Fig.~\ref{adqh} and in accordance with
the above discussion, $\sigma_{22}$ is dominant and completely
defines the backward scattering (note the scale factor 0.2 in the
figure). The $\sigma_{11}(\Theta)$ and $\sigma_{33}(\Theta)$
angular distributions are forward peaked but the asymmetry is much
more pronounced in the $s$-channel. The sum of interference terms
results in the distribution with a broad maximum at
$cos\Theta\sim$0.

The presented results are the same for the $u\pi^-\to u\pi^-$ and
$d\pi^+\to d\pi^+$ reactions because their flavor factors are the
same (see Table II). For other channels of the quark-pion
scattering, the shape of angular distributions is similar to that
for $u\pi^-\to u\pi^-$ but the magnitude is lower in accordance
with the flavor factors in Table II. Since a pion consists of a
quark-antiquark pair, the substitution $u\to d$ for quarks does
not change the calculated scattering cross sections results in
Fig.~\ref{crosqh} and angular distributions in the $s$- and
$u$-channels shown in Fig.~\ref{adqh} are exchanged.

\section{Concluding remarks}

In-medium elastic scattering of quarks is evaluated by treating
quarks in terms of the two-flavor PNJL model  accounting for their
scalar and pseudoscalar interactions  and  using  the large
$N_c$ approximation. The model parameters are in
agreement with low-energy static mesonic properties and lattice
QCD results for gluons. General trends of the energy and
temperature dependence of $\sigma_{el}$ and
$\d\sigma_{el}/dcos\Theta$ are investigated in a large range of
$\sqrt{s}$ and $T$ at both below and above the Mott temperature.
Essential influence of the flavor of interacting quarks  on the
scattering results is noted. The comparison with the earlier works
shows that the $SU_f(2)$ PNJL results agree with those of the NJL
model for the same $SU_f(2)$ symmetry (at $T>T_{Mott}$) but
noticeably below in the case of three flavors. This difference is
due to a larger number of contributing intermediate states for the
quark scattering in the case of $SU_f(3)$ symmetry.

The above PNJL results are presented for the vanishing chemical
potential. However, $\mu$ enters not only into the Pauli exclusion
factor but also into the polarization function $\Pi(k^2)$ which
makes masses and coupling constants to be $\mu$-dependent. To
consider the quark-baryon case, an additional issue arises: one
should treat properly the loop formed by the quark and the diquark
that form the baryon. As demonstrated in ~\cite{Bl12}, taking into
account the finite chemical potential results in some suppression
of the scattering cross sections at large $\mu$. The PNJL approach
should also be generalized to the $SU_f(3)$ symmetry. This allows
one to extend the set of reactions including strange quark and
strange hadrons and  to effect  non-strange reactions due to
increasing a number of possible intermediate states, as noted
above. This work is in progress now.

The first predictions for the  quark-pion scattering are given.
The integral cross sections for this reaction are  higher than
those in the quark-quark case.  The quark differential $cos
\Theta$-distribution is practically constant with some enhancement
in the backward direction at $\sqrt{s}\gsim$1 GeV. The developed
technique and obtained results can be applied in the kinetic
approach~\cite{AMPT,PHSD,MA12} to take into account the interaction
between constituents of quark-gluon and hadronic phases. The
derived formulae  for the elastic quark-hadron scattering can be
easily generalized to the expression for gamma and dielectron
emission~\cite{LCMBK12}. This new channel  can definitely give
rise to an observable effect and would confirm the existence of
the quark-hadron interaction.

Another possible implementation of these results is a study of
transport properties of the system. For example, the shear
viscosity $\eta$ can be estimated in the so-called relaxation time
approximation  by using the average momentum loss $\bar{p}$, the
quark densities $n_i$ and the mean life time as $\displaystyle
\eta \sim \sum_i n_i \bar{p}_i /\lambda_i$, the mean life time
being inversely proportional to the quark cross sections
$\displaystyle \lambda^{-1}=\sum_i n_i \sigma_{qi}$
Ref.~\cite{KTV10}. In the two-flavor NJL model, this shear
viscosity was estimated in Ref.~\cite{ZHKN95} for the chiral limit
$m=0$. One should not expect a noticeable difference with our
model treating a quark system with the finite quark masses in the
$SU_f(2)$ PNJL model, especially at high temperature where the NJL
model is more justified. However, for the mixed quark-hadron
system considered  in~\cite{AMPT,PHSD,MA12}  the shear viscosity
should be smaller, because the $qH$ elastic cross section is
noticeably larger than the quark-quark one and particle density of
quarks and pions is rather abounded in the mixed phase. Certainly,
this problem deserves a more elaborated study, for example,
by the method developed in Ref.~\cite{Oz12} .\\[2mm]

{\bf Acknowledgments}\\
 We would like to thank H.~Berrehrah, R.~Marty and V.~Priezzhev
for fruitful discussions and constructive remarks. We are grateful
to E.~Bratkovskaya, W.~Cassing and O.~Linnyk for their continuous
interest in this work. This work was supported in part (Yu. K.) by
RFFI grants 13-01-00060, 12-01-00396.

 \end{document}